%% file: ms.tex
\documentclass[fleqn,usenatbib]{mnras}

\usepackage{newtxtext}

\usepackage[T1]{fontenc}
\usepackage{ae,aecompl}

\usepackage{graphicx}	
\usepackage{amssymb}	
\usepackage{subcaption}	
\usepackage{float}
\usepackage{stfloats}   
\usepackage{units}
\usepackage{bm}
\usepackage{multirow}

\usepackage{mathtools}	

\newcommand{\lcdm}{$\Lambda$CDM~}
\newcommand{\hMpc}{$h^{-1}$Mpc}

\title[Cosmic web homology for \lcdm cosmologies]{Persistent homology of the cosmic web. I: \\ Hierarchical topology in $\bm{\Lambda}$CDM cosmologies}

\author[G. Wilding et al.]{Georg Wilding$^{1,2,3}$\thanks{E-mail: wilding@astro.rug.nl},
Keimpe Nevenzeel$^{1}$,
Rien van de Weygaert$^{1,3}$,
Gert Vegter$^{2,3}$,
\newauthor
Pratyush Pranav$^{1,4,5}$,
Bernard J. T. Jones$^{1}$,
Konstantinos Efstathiou$^{6,2,3}$,
\newauthor
Job Feldbrugge$^{7,8}$
\\
$^{1}$Kapteyn Astronomical Institute, University of Groningen, PO Box 800, 9700 AV Groningen, The Netherlands\\
$^{2}$Bernoulli Institute for Mathematics, Computer Science and Artificial Intelligence, University of Groningen,\\\quad PO Box 800, 9700 AV Groningen, The Netherlands\\
$^{3}$Centre for Data Science and Systems Complexity, University of Groningen, PO Box 800, 9700 AV Groningen, The Netherlands\\
$^{4}$Univ Lyon, ENS de Lyon, CNRS, Centre de Recherche Astrophysique de Lyon UMR5574, FV69007, Lyon, France\\
$^{5}$Technion -- Israel Institute of Technology, Haifa, 32000, Israel\\
$^{6}$Division of Natural and Applied Sciences and Zu Chongzhi Center for Mathematics and Computational Science,\\\quad Duke Kunshan University, No. 8 Duke Avenue, Kunshan 215316, Jiangsu Province, China\\
$^{7}$Perimeter Institute, 31 Caroline St N, Waterloo, ON N2L 2Y5, Canada\\
$^{8}$Department of Physics, Carnegie Mellon University, 5000 Forbes Ave, Pittsburgh, PA 15217, USA
}

\date{Accepted XXX. Received YYY; in original form ZZZ}

\pubyear{2021}

\begin{document}
\label{firstpage}
\pagerange{\pageref{firstpage}--\pageref{lastpage}}
\maketitle

\begin{abstract}
Using a set of $\Lambda$CDM simulations of cosmic structure formation, we study the evolving connectivity and changing topological structure of the cosmic web using state-of-the-art tools of multiscale topological data analysis (TDA). We follow the development of the cosmic web topology in terms of the evolution of Betti number curves and feature persistence diagrams of the three (topological) classes of structural features: matter concentrations, filaments and tunnels, and voids. The Betti curves specify the prominence of features as a function of density level, and their evolution with cosmic epoch reflects the changing network connections between these structural features.

The persistence diagrams quantify the longevity and stability of topological features. In this study we establish, for the first time, the link between persistence diagrams, the features they show, and the gravitationally driven cosmic structure formation process. By following the diagrams' development over cosmic time, the link between the multiscale topology of the cosmic web and the hierarchical buildup of cosmic structure is established.

The sharp apexes in the diagrams are intimately related to key transitions in the structure formation process. The apex in the matter concentration diagrams coincides with the density level at which, typically, they detach from the Hubble expansion and begin to collapse. At that level many individual islands merge to form the network of the cosmic web and a large number of filaments and tunnels emerge to establish its connecting bridges. The location trends of the apex possess a self-similar character that can be related to the cosmic web's hierarchical buildup. We find that persistence diagrams provide a significantly higher and more profound level of information on the structure formation process than more global summary statistics like Euler characteristic or Betti numbers.
\end{abstract}





\input{1_introduction}
\input{2_simulation}
\input{3_results1}
\input{4_results2}
\input{5_summary}


\section*{Acknowledgements}\label{sec:acknowledgements}

This project is conducted in part at the Centre for Data Science and Systems Complexity at the University of Groningen and is sponsored with a Marie Sk\l odowska-Curie COFUND grant, no. 754315.
Research at Perimeter Institute is supported in part by the Government of Canada through the Department of Innovation, Science and Economic Development Canada and by the Province of Ontario through the Ministry of Colleges and Universities.
PP acknowledges support from the European Research Council (ERC)
under the European Union's Horizon 2020 research and innovation programme (grant agreement ERC advanced grant 740021 -- Advances in the Research on THeories of the dark UniverSe (ARTHUS), PI: Thomas Buchert).
We wish to thank Michael Kerber for his help with respect to the computational topology codes that allowed our calculations,
Patrick Bos for providing the Gadget simulation data and Raul Bermejo for helpful discussions and suggestions.

\section*{Data availability}
The N-body simulation data analysed in this study were provided by Patrick Bos and will be shared with his permission.




\bibliographystyle{mnras}
\bibliography{literature} 


\bsp	
\label{lastpage}
\end{document}

%% file: 1_introduction.tex
\section{Introduction}
In this study we analyse the topological structure and connectivity of the cosmic web~\citep{bond1996filaments,weygaert2008clusters} in terms of the multiscale topological formalism of persistence and Betti numbers. These state-of-the-art tools of topological data analysis (TDA) represent measures of structural aspects of the cosmic web~\citep{weygaert2011alpha,sousbie2011persistent, nevenzeel2013triangulating, shivashankar2016felix, pranav2017topology,xu2019finding,biagetti2020}. With a solid mathematical foundation in the context of algebraic and computational topology~\citep{edelsbrunner2010computational}, they offer an intricate quantitative description of how the structural components of the cosmic web are assembled and organised within its complex network. The principal intentions of the present study are (1) to assess and quantify the connectivity of the cosmic web in terms of the levels at which its various structural components get joined into the overall weblike network, (2) establish the relationship between the characteristics of the Betti number curves and persistence diagrams and the gravitationally driven cosmic structure formation process, (3) to explore the sensitivity of the structure and topology of the cosmic web to the underlying cosmology and (4) to assess the extent to which the topological measures are able to extract cosmological information. This concerns aspects such as the nature of dark matter, dark energy, possible deviations from standard gravity, and/or non-Gaussian initial conditions.

The use of persistence diagrams as a tool of topological analysis will prove valuable, as it enables us to measure non-linear features in the large-scale structure. In line with using it to differentiate between cosmologies, we aim to turn this manner of analysing persistence into a new probe for fundamental cosmology and physics in general. Ultimately, we will apply this probe also to observational data, with the aim of differentiating between models and providing constraints on the nature of dark matter, dark energy and other global cosmologically relevant factors.

\subsection{Cosmic web: Connectivity}
On Megaparsec scales the matter and galaxy distribution defines an intricate multiscale inter-connected network, known as the \textit{cosmic web}~\citep{bond1996filaments}. It represents the fundamental spatial organisation of matter on scales of a few up to a hundred Megaparsec. Galaxies, intergalactic gas and dark matter arrange themselves in a salient, wispy pattern of dense compact clusters, long elongated filaments, and sheetlike tenuous walls surrounding near-empty void regions. Maps of the nearby cosmos produced by large galaxy redshift surveys such as the 2dFGRS, the SDSS, and the 2MASS redshift surveys~\citep{colless2003,tegmark2004cosmological,huchra20122mass}, as well as by recently produced maps of the galaxy distribution at larger cosmic depths such as VIPERS~\citep{vipers2013} and GAMA~\citep{gama2009}, have revealed the existence of this structure. Filaments are the most visually outstanding features of the Megaparsec universe, in which around $50\%$ of the mass and galaxies in the universe reside. On the other hand, almost 80\% of the cosmic volume belongs to the interior of voids~\citep[see e.g.][]{cautun2014,ganeshaiah2018cosmicballet}. Together, they define a complex spatial pattern of intricately connected structures, displaying a rich geometry with multiple morphologies and shapes. This complexity is considerably enhanced by its intrinsic multiscale nature, including objects over a considerable range of spatial scales and densities. For a recent up-to-date report on a wide range of relevant aspects of the cosmic web, we refer to the volume by~\cite{iau308}.

The organization of this network in an ordered web -- in which voids are surrounded by walls and filaments, connecting at high-density compact clusters at the nodes evidently -- is a characteristic that is in need of a systematic and quantifiable characterisation. Filaments appear at the edges of the walls in the mass distribution. The way in which the various features connect into the weblike pattern pervading space includes local as well as global aspects. Locally, it concerns questions like the dependence of the number of connecting filaments on the properties of a (cluster) node, or the connection between walls and surrounding or embedding filaments. Globally, it pertains to issues of percolation, i.e. how fast and at what level the various structural elements are connecting up in a network that permeates an entire volume.

The study by~\citet{aragon2010multiscale} was amongst the first to address this question systematically, and established that the number  of connecting filaments is linearly increasing with the mass of the node and is typically in the order of 3 to 5 filaments per node. Recent work by~\citet{codis2018connectivity} on the basis of a topological analysis has confirmed this trend.

The more global aspect of connectedness concerns the overall percolation properties of the weblike network, focusing on how the various structural features connect up into the final permeating network. Early studies within the context of percolation theory by Zeldovich and coworkers~\citep{zeldovich1982,shandarin1983percolation,klypin1993percolation,colombi2000tree}, and others~\citep{dekelwest1985,sahni1997}, explored the spatial connectedness of galaxies as a function of linking length, assessing the length at which all galaxies would link up and comparing this with the expectation for different cosmologies. For the connectedness of the structural components of the cosmic web -- nodes, filaments, walls, and voids -- a similar approach may be pursued by using the criterion or physical quantity according to which they are identified.

In the present study we restrict ourselves to using the density field for identification of structures affiliated to the cosmic web. The levels over which filaments and walls exist in the density field establishes the connection of the different components. By following the changing pattern and population of components at different density levels, one may study how the structural elements have connected into a volume pervading network. Rather than using density, a more sophisticated analysis would use a physical influence that is more relevant for distinguishing cosmic web identities. An example of this is the tidal force field or the closely related deformation field. The recent analytical formulation of the caustic skeleton of the cosmic web on the basis of the eigenvalues and eigenvectors of the deformation field~\citep{feldbrugge2019} will therefore yield a more detailed and profound quantitative characterisation of the global cosmic web connectedness.

Following this procedure defines a sophisticated multiscale analysis of the connectivity of the cosmic web. The mathematical formalism for this we find in topology, more specifically within homology theory.

\subsection{Topology: Betti numbers and persistence}
Topology is the branch of mathematics that addresses the connectivity of this multitude of features, as well as their occurrence in various dimensions and shapes. Early cosmological studies that studied the topology of the cosmic mass distribution restricted themselves to the evaluation of the genus and Euler characteristic of the cosmic mass distribution for the corresponding iso-density surfaces. Gott and collaborators~\citep{gott1986sponge,hamilton1986} studied the genus as a function of density threshold. Later, more discriminative topological information became available with the introduction of Minkowski functionals~\citep{mecke1993robust,schmalzing1997}. However, nearly without exception these studies had a largely heuristic character and focused on global statistical assessments of the cosmic mass distribution. The first study focusing on the connectivity of distinct morphological elements in the mass distribution is the SURFGEN formalism developed by~\cite{sahni1998}. It uses Minkowski functionals to define \textit{shapefinders}, allowing the identification of morphological features of different geometric shapes, and carry out a systematic assessment of their embedding  within the overall cosmic mass distribution~\citep{sheth2003,shandarin2004,sheth2005}.

Van de Weygaert and collaborators~\citep{weygaert2010alphashapetopology,weygaert2011alpha} introduced the concept of homology, Betti numbers~\citep{poincare1892betti} and persistence~\citep{edelsbrunner2002topological,edelsbrunner2010computational}, in a cosmological context. These are homology measures, concepts of algebraic and computational topology, describing in a quantitative manner how features in a manifold are connected through their boundaries~\citep{munkres1984}. These early studies assessed  Betti number systematics in a range of weblike spatial mass and galaxy distributions, for which they provide a summary of information on the topology of the cosmic mass distribution. This was followed up by recent studies that invoked homology in a cosmological context along more systematic and formalised lines~\citep{weygaert2011alpha,sousbie2011persistent,park2013,pranav2017topology,pranav2019topology,feldbrugge2019}.

{\it Betti numbers} are topological invariants that formalise the topological information content of the cosmic mass distribution in terms of the population of topological features~\citep{edelsbrunner1994,zomorodian2005,robins2006,edelsbrunner2010computational,wasserman2018}. The zeroth Betti number counts the number of connected components, the first Betti number is the number of independent loops, while the second Betti number is the number of independent shells enclosing troughs. Within the context of the spatial pattern of the cosmic web, tunnels are intimately related to loops of filamentary bridges of the cosmic web connecting the overdense clusters. It is important to appreciate that the homological measures are \textit{fundamentally non-local}. While homology and the Betti numbers do not fully quantify the topology of a manifold, they extend the information beyond conventional cosmological studies of topology in terms of genus and Euler characteristics.

The profound significance of Betti numbers is underlined by their intimate relationship to the singularity structure of the cosmic density field~\citep{morse1925, milnor1963}. According to Morse theory the topology of a field is coupled to the presence, location and nature of the singularities. It reflects the notion that the topology of a manifold changes once a singularity is added, or removed, upon variation of the level set. As a result, the existence of and connectivity between topological features is completely determined by the location and nature of the critical points in a density field. The importance and prominence of topological features is characterized through their persistence~\citep{edelsbrunner2002topological,edelsbrunner2010computational}.

Persistence facilitates the assessment of the multiscale nature of the topology of the Megaparsec cosmic mass distribution. Of key significance is the ability to assess its structural nested hierarchy, i.e., the possibility to study how the structural elements of the weblike network connect up upon variation of the level set. The corresponding change in topology represents a highly informative and versatile description of the connectivity of the cosmic web network~\citep{edelsbrunner2002topological,edelsbrunner2010computational}. Persistence relates the \textit{creation} or \textit{birth} of topological features (e.g. holes) that constitute the mass distribution with that of their \textit{annihilation} or \textit{death} upon variation of the level set.

\subsection{This study: persistent topology of the cosmic web}
Recent years have seen a proliferation of scientific studies invoking persistent topology to characterize the complexity of a large diversity of systems and processes (see~\citealp{wasserman2018} for a recent review), ranging from brain research~\citep{petri2014,reimann2017}, materials science~\citep{hiraoka2016} to astrophysics and cosmology.~\citet{sousbie2011persistent},~\citet{sousbie2011persistent2},~\citet{shivashankar2016felix}, and~\citet{pranav2017topology} invoke persistence with the purpose to characterize the spine of the cosmic web~\citep{bond1996filaments,weygaert2008clusters,aragon2010multiscale,cautun2014,libeskind2018} and its connectivity structure. Recently,~\citet{kimura2016} determined persistence diagrams for (small) volume-limited samples of the DR12 release of the SDSS galaxy redshift survey in an attempt to characterize the topology of the spatial galaxy distribution, while~\citet{xu2019finding} used persistence to identify voids and filaments in heuristic models of the cosmic matter distribution~\citep[also see][]{shivashankar2016felix}.~\citet{kono2020} applied topological data analysis towards studying baryonic acoustic oscillations in the galaxy distribution, while~\cite{biagetti2020} studied persistence properties of the large scale matter distribution in cosmologies with non-Gaussian primordial conditions~\citep[also see][]{feldbrugge2019}. The explicit application of homology measures in the study of the primordial temperature perturbations in the cosmic microwave background are reported in~\citet{pranav2019unexpected} and~\citet{adler2017}. 

At a more fundamental level,~\citet{codis2018connectivity} based their assessment of the connectivity of the nodes of the cosmic web on the persistent characterisation of the cosmic web's spine~\citep[also see][]{aragon2010spine}. The concepts of persistence and Betti numbers also offer a natural means of following the evolving topology of the reionization bubble network~\citep{elbers2019}. In another astrophysical context, they were used to describe the topological structure of interstellar magnetic fields~\citep{makarenko2018}.

\bigskip
Following the work laid out in~\citet{weygaert2011alpha},~\citet{nevenzeel2013triangulating},~\citet{pranav2017topology},~\citet{pranav2019unexpected},~\citet{pranav2019topology} and~\citet{feldbrugge2019}, in the present study we extend the topological analysis of the cosmic web to the analysis of the redshift evolution of structure on simulations within the \lcdm cosmology. In Section~\ref{sec:methods} we first describe the simulation of structure formation in \lcdm cosmology that we used in the present study, as well as the tools, methods and implementation of persistent topology. The Betti numbers and persistence of the dark matter distribution at redshift $z=0$ is discussed in Section~\ref{sec:resultsz0}, with the purpose of identifying the topological characteristics of the weblike mass distribution. The systematic development of these characteristics in the evolving mass distribution in \lcdm cosmologies follows in Section~\ref{sec:resultsevolution}. We conclude with the summary and conclusions in Section~\ref{sec:summary}.

%% file: 2_simulation.tex
\section[Simulations, Tools \& Methods]{Simulations, Tools \& Methods}\label{sec:methods}
Our analysis concentrates on the dark matter distribution in a $\Lambda$CDM cosmology. The gas, halo and galaxy distribution in this cosmology possess similar topological characteristics, although the details display significant and systematic differences. We will address the topological characteristic of, for example, the dark matter halo distribution in accompanying studies~\citep[see e.g.][]{bermejo2020}.

\begin{figure*}
 \centering
\includegraphics[width=\textwidth]{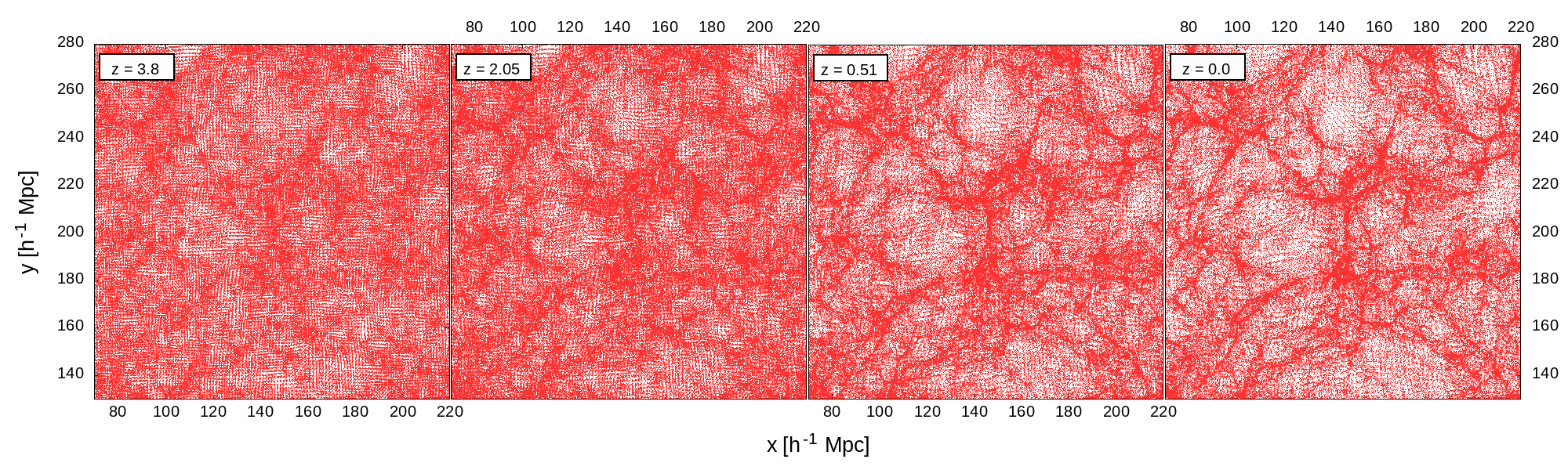}
 \caption{\textbf{Cosmic web evolution in $\bm{\Lambda}$CDM cosmology.} We show the evolution at four redshifts, starting at $z = 3.8$ at the left and proceeding clockwise. The slices show a 150 by \unit[150]{\hMpc{}} as a projection from a \unit[24]{\hMpc{}} thick region around a height of \unit[117]{\hMpc{}}.}\label{fig:1_gadgetdtfe}
\end{figure*}

\subsection{Simulation and density field}
We analyse the simulated evolving dark matter distribution in a set of \lcdm simulations of cosmic structure formation. The simulations were performed with Gadget 3~\citep{dolag2004numerical,springel2005cosmological}. We use five runs, each with 256$^3$ particles of mass $0.443\cdot 10^{10} h^{-1}M_\odot$ in a box of 300$h^{-1}$ Mpc, using periodic boundary conditions. The cosmological parameters are based on the WMAP3 data~\citep[see][for a detailed discussion]{bos2012darkness}.

The dark matter particle distribution produced by the Gadget simulations is transformed into a density field by means of the Delaunay Tessellation Field Estimator (DTFE)~\citep{schaap2000continuous,weygaert2008cosmic,cautun2011dtfe}. To this end, the Delaunay tessellation~\citep{delone1934sphere,okabe2000} of the N-body particle distribution is determined, and the densities at each vertex of the tessellation computed from the inverse of the volume of the star of the vertex, the union of all Delaunay tetrahedra incident to the vertex. The densities at the vertices (which correspond to the particles in the simulation) are then linearly interpolated to a regular grid. By using the density and shape adaptive properties of the Delaunay tessellation~\citep[see][]{weygaert2008cosmic}, DTFE optimally retains the multiscale, geometric and topological nature of the underlying mass distribution that the N-body particle distribution is supposed to sample. The density values are specified in terms of the density contrast $\delta(\textbf{x},t)$ 
\begin{equation}\label{eq:delta}
    \delta(\textbf{x},t) = \frac{\rho(\textbf{x},t) - \rho_u(t)}{\rho_u(t)} ,
\end{equation}
with $\rho$ the densities from the DTFE, and $\rho_u(t)$ the global density value at the appropriate cosmic epoch. As $\delta$ ranges from $-1$ to $\infty$, in our plots we usually use $\delta + 1$, in order to enable logarithmic scale plots (by avoiding negative values).

For the topological analysis, we use 8 different snapshots of the simulation. These correspond to the redshifts $z = 3.8$, $2.98$, $2.05$, $1.00$, $0.51$, $0.25$, $0.1$ and $0.00$. To get an impression of the resulting spatial pattern in the matter distribution, Fig.~\ref{fig:1_gadgetdtfe} shows the particle distribution in a $300 \times 300 \times 24$ \hMpc~slice around a height of \unit[117]{\hMpc}. The evolution of the web-like structure is followed through four snapshots, from $z=3.8$ down to the current epoch at $z=0$. The four panels show how the relatively low contrast mass distribution at high redshift evolves in the prominent and complex web-like pattern that pervades the entire box and attains scales in the order of dozens of Megaparsec.

\subsection{Cosmic web evolution \& topology}
\label{sec:cosmicwebevolve00}

\subsubsection{Density field dynamics}
At all snapshots we see the weblike pattern characteristic of the quasi-linear mass distribution that evolves from the initial linear gravitational growth to more advanced non-linear stages~\citep{bond1996filaments,weygaert2008clusters,aragon2010multiscale,cautun2014}. The set of panels reveal how gravitational contraction and collapse manifests itself into increasing density contrasts and gradual contraction of overdensities into more compact clump-like, filamentary and wall-like features, and ever emptier void regions.

The hierarchical buildup of structure in the $\Lambda$CDM scenarios involves the emergence of ever larger complexes or islands, the hierarchical development of large near-empty void regions that emanate from the merging of smaller scale troughs~\citep[see][]{shethwey2004,aragon2013} and the establishment of major filamentary arteries as the transport channels along which mass flows through the universe, connecting all mass concentrations throughout it. We first see the emergence of weblike structures at small scales, which through gravitational interactions subsequently grow and merge into larger structures. While this happens, the evolution of structure also establishes new or more pronounced connections. Towards the current cosmic epoch at $z=0$, it yields the characteristic web-like pattern dominated by filaments and voids on scales of tens to even hundreds of Megaparsec.

The left-most panel in Fig.~\ref{fig:1_gadgetdtfe} shows the mild density contrast at a redshift $z=3.8$. By $z=2.05$, we see that the mass distribution has evolved into one marked by a substantially higher density contrast. The mild density enhancements at $z=3.8$ have contracted into steep density ridges and complexes, within which we observe compact clumps of high density and moderately dense elongated filaments. These island complexes appear to be connected by lower contrast filamentary and wall-like bridges. We see that the regions of lower density have grown in size and contrast, into large near-empty troughs. It is the result of the continuation of gravitational contraction and collapse, manifesting itself into increasing density contrasts and gradual contraction of overdensities into more compact clump-like, filamentary and wall-like features, and ever emptier void regions.

\begin{figure*}
 \centering
 \includegraphics[width=\textwidth]{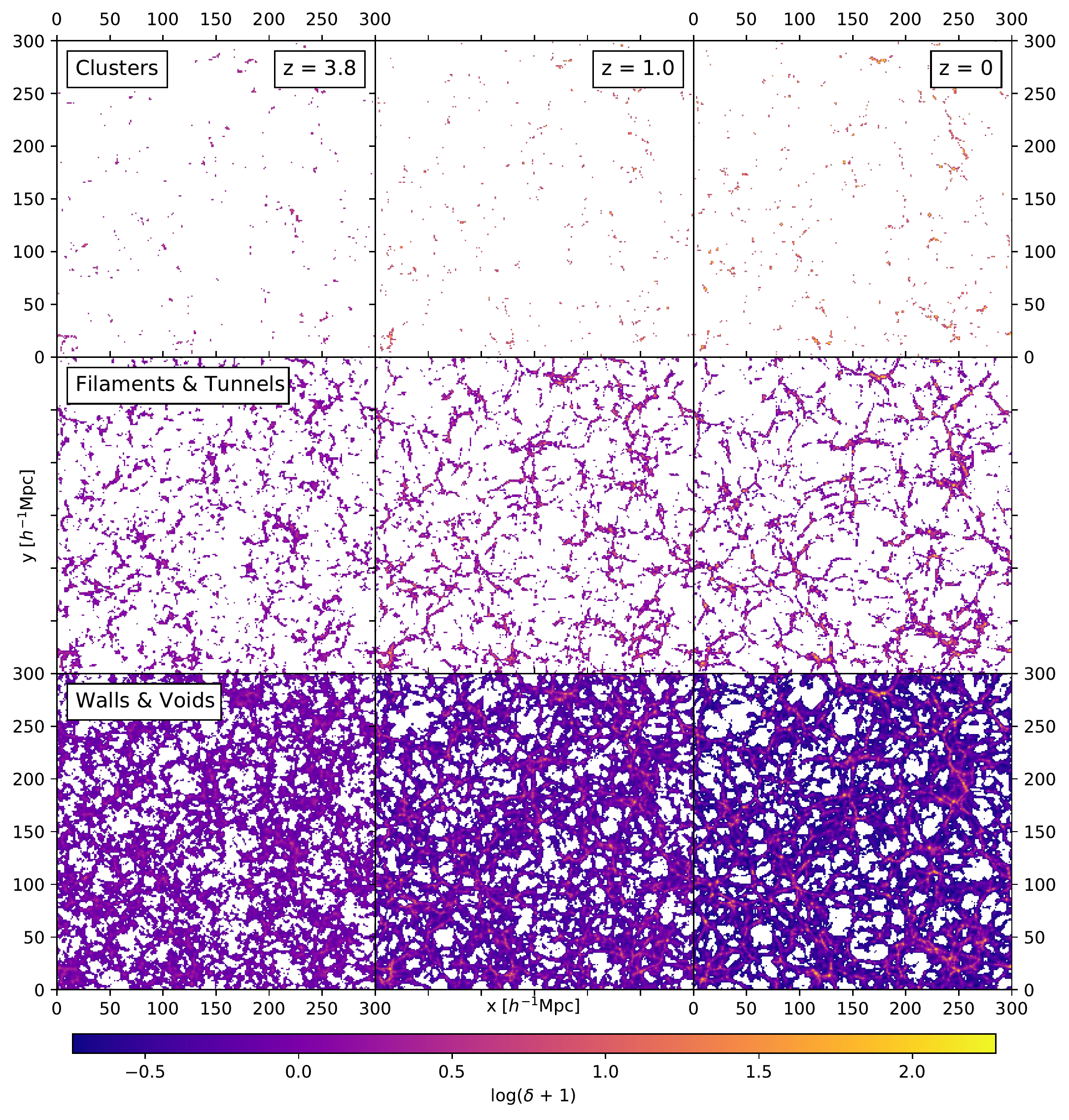}
 \caption[Evolution of structure.]{\textbf{Density superlevel sets of the $\bm{\Lambda}$CDM mass distribution.} Evolution of structure for three redshifts ($z = 3.8$, $1$ and $0$, from left to right), in a \unit[24]{\hMpc{}} slice around a height of \unit[117]{\hMpc{}}. The structure is depicted as superlevel sets with three different thresholds to outline the disjoint nodes of the cosmic web (top row), its filamentary structure (middle row) and the walls enclosing the cosmic voids (for the calculation of the thresholds see Section~\ref{sec:densityvisualisation}). The evolution of structure and the emergence of the increasingly geometric and organised web is particularly visible in the middle row, where noisy, short and disjoint (but already elongated) clumps connect up to form a network of long and more massive filaments, that (in three dimensions) fills the whole volume.}\label{fig:2_dtfe_level}
\end{figure*}

\subsubsection{The topological point of view}
For a visual appreciation of the effects of these dynamical and hierarchical processes on the changing topology of the cosmic mass distribution Fig.~\ref{fig:2_dtfe_level} follows the cosmic web patterns at three different structural levels. The figure shows these patterns in terms of the density superlevel sets at three density thresholds, and follows their evolution at three redshifts, $z=3.8$, $z=1.0$ and $z=0$. The three threshold levels have been carefully chosen such that the superlevel sets are typically representing the presence for three structural components of the cosmic web (see Section~\ref{sec:bettivisual} for their definition). An immediate visual impression of the evolving structure from redshift $z = 3.8$ to $0$ is the increasing sharpness of the morphological features in the mass distribution. It is most outstanding in the development of the intricate filamentary network (middle row) and the pronounced topology marked by void cavities (bottom row). 

The top row shows the structures at the highest threshold, at which level we observe the presence of high-density peaks and islands -- their immediate surroundings -- which congregate near the nodes of the cosmic web. Following their evolution, from top left to top right, we observe two processes. Existing peaks and islands merge into higher density compact clumps. Also, we see the emergence of new peaks and islands that have gravitationally grown over the threshold level. The latter occurs abundantly from $z=1.0$ to $z=0$, to such an extent that at $z=0$ we start to see that the clumps delineate large elongated features, the superclusters that trace the most prominent filaments and walls of the cosmic web.

At the intermediate level, the superlevel pattern is shown in the central row of Fig.~\ref{fig:2_dtfe_level}. At this level, filaments and walls -- and the tunnels that go along with them -- manifest themselves as the dominant structure visible. Going from $z=3.8$ to $z=0$, we also note that these features are generally smaller at the earlier epochs, and that we see them connect up into ever larger and more massive features and agglomerates. It demonstrates the hierarchical buildup of the filamentary and wall-like backbone of the cosmic web~\citep[see][]{cautun2014}. It is also interesting that the features at $z=0$ are more sharply outlined than their peers at $z=3.8$, which are shorter and stubby, as a result of their gravitational contraction into more pronounced and compact configurations.

At the lowest threshold, represented by the panels in the bottom row, nearly all structure has percolated into a foam-like network that permeates the entire cosmic volume. This is certainly the case for the $z=0.0$ cosmic web, while at earlier epochs we still find disconnected parts: at the smoothing scale of the density field, the universe is not yet permeated by a percolating cosmic web. Also the void population is evolving characteristically, from a large number of smallish underdense regions at $z=3.8$, to one of a considerably lower number of much larger void regions.  It illustrates the hierarchical nature of void evolution, akin to a soapsud of bubbles which merge into ever larger ones~\citep[see][]{shethwey2004,aragon2013}.

The final pattern and topology of the resulting hierarchically evolving mass distribution is determined by the relative dynamical timescales at the different spatial scales of the mass distribution. For the Gaussian initial conditions in the early Universe, this is fully determined by the primordial power spectrum of density and velocity fluctuations. Processes in the early Universe, as well as important factors such as the nature of dark matter, arrange the power spectrum. It therefore determines in how far we are dealing with a clumpy distribution of objects arranged in larger-scale web-like configurations, or one in which the structures on the scale dominating at that epoch have a more coherent appearance. The connectivity of these patterns will be radically different. It translates into fundamental differences in the multiscale -- and hence persistent -- topology, representing the global phenomenon of connectivity that cannot be described by power spectra or correlation functions.

The present study is based on the realisation that the visually appreciable change in multiscale topology as we proceed from the panel in Fig.~\ref{fig:1_gadgetdtfe} at $z=3.8$ up to the panel at the current cosmic epoch at $z=0$ should allow us to determine with considerable precision the underlying cosmology.

\subsection{Persistent homology: background \& implementation}
It is useful to summarise the terminology relevant to this study. Technical details can be found in many of the previously cited papers~\citep[e.g.]{edelsbrunner2002topological,edelsbrunner2010computational,wasserman2018}, while a more detailed summary than can be given here is to be found in~\citet{pranav2017topology}.

When describing the structural elements of the cosmic web, we loosely talk in terms of `clusters, filaments, and voids'. In topology we speak descriptively of `islands, loops, and shells' or of `components, tunnels, and cavities'. More precisely, these structures are referred to as $k$-cycles: $0$-cycles (a connected component), $1$-cycles (loops surrounding tunnels) and $2$-cycles (shells enclosing voids). Formally defined in terms of \emph{homology groups}, the number of independent structures, and the size of these groups, are the \emph{Betti numbers} $\beta_k$. The topology of structures in three dimension is characterized through a triple of Betti numbers: $\beta_0$, $\beta_1$, $\beta_2$.

At any instant in a cosmological simulation, the character of the topology of the superlevel density field (outlined by structures above a threshold density) changes with the value of the threshold (see Fig.~\ref{fig:2_dtfe_level}). With the topology tied to the three Betti numbers, we will obtain three curves determining the Betti numbers as a function of the threshold. The curves will vary with cosmic epoch, and at each characterize the structure. Topology addresses the identity and shape of each superlevel set and the spatial connectivity of features like islands, tunnels, and cavities or voids. Islands in a superlevel set are the regions with a mass density in excess of a specific threshold. One may study the connections with different thresholds, and with decreasing threshold determine how many tunnels percolate their interior, compute the number of cavities they encompass, and consider a range of additional questions of interest (e.g. the shape or orientation of either of the relevant components). One of the most important notions in this context is the fundamentally non-local character of the topological measures.

\begin{figure}
	\centering
 \includegraphics[width=\columnwidth]{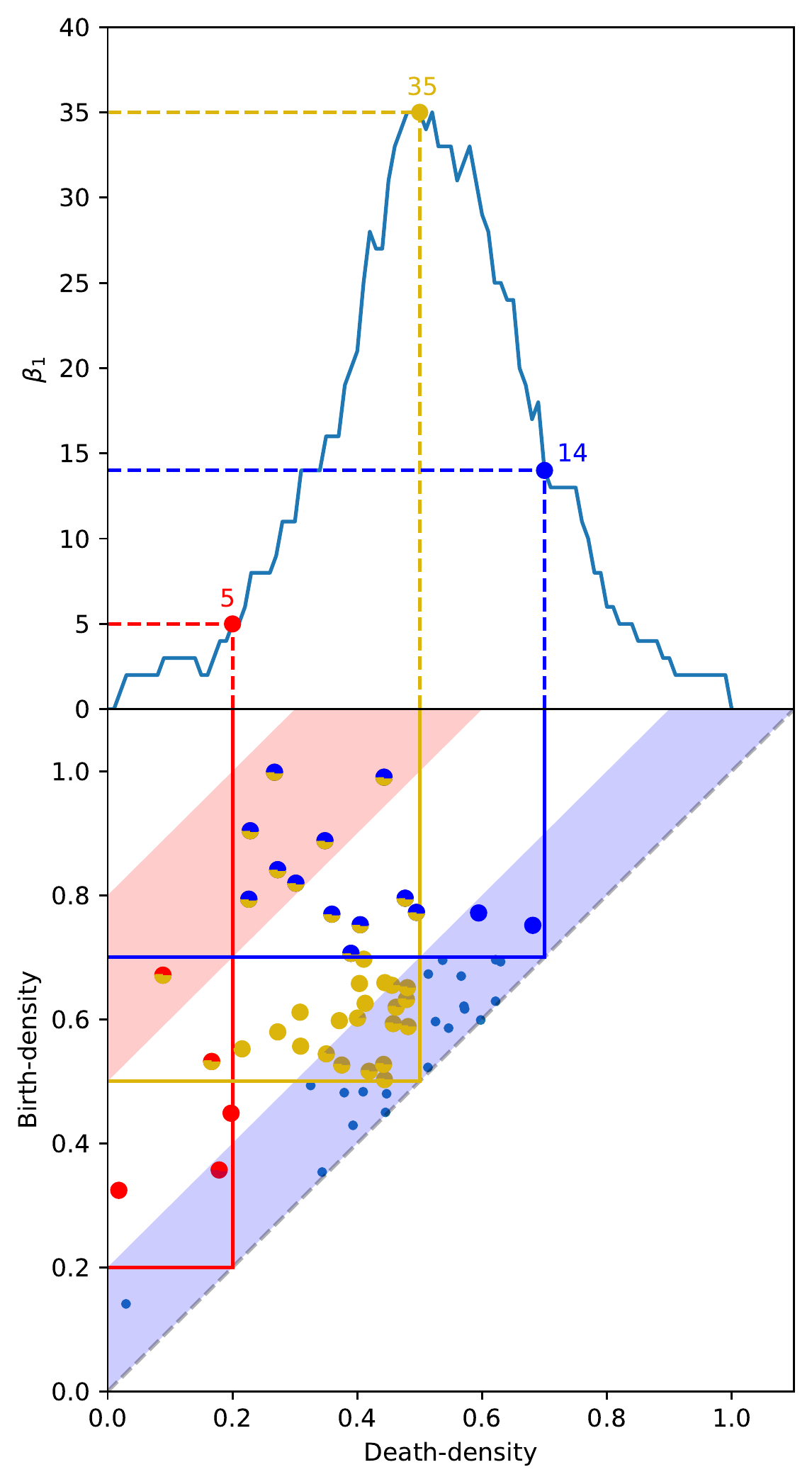}
    \caption{\textbf{A mock persistence diagram and Betti curve.} In the lower panel we show 60 randomly generated persistence points (blue dots), with
      birth- and death-densities between between 0 and 1. The points signify $k$-dimensional features that are respectively born and destroyed at
      these densities. All points lie above the diagonal (grey dashed line). Points with low
      persistence (i.e., horizontal distance to the diagonal) have a small distance to the diagonal, indicated by the blue shaded region adjacent to the diagonal. The red shaded region in the upper part of the panel covers an area
      with several points of high persistence. The Betti curve in the top panel can be calculated from the diagram below by ``sliding'' a rectangular region along the diagonal and counting the persistence points that lie in it. This is shown for three densities: 0.2 with 5 features (in red), 0.5 with 35 features (in yellow) and 0.7 with 14 features (in blue).}
    \label{fig:3_sample_persistence}
\end{figure}

On a more detailed level than that of the Betti curves, and focusing in particular on the multiscale nature and interactions of these features, we can depict them in so-called persistence diagrams. In these diagrams features are represented as points in two dimensions, with the coordinates being the threshold values at which they appear in the superlevel density field (they are \emph{born}) and at which they disappear again (they \emph{die}). Accordingly, these values of the threshold density are referred to as \emph{birth}- and \emph{death}-density. Persistence diagrams show features at all densities, as opposed to Betti curves which show only features ``alive'' at specific densities. The relation between a Betti curve and a persistence diagram is outlined in Fig.~\ref{fig:3_sample_persistence}. We show a mock persistence diagram (bottom panel) and the straightforward connection to the corresponding Betti curve (top panel). At each density, the Betti curve shows the number of existing features, i.e., features that have been born before (at higher birth-densities) and will die later (at lower death-densities)\footnote{This assumes a decreasing threshold of the superlevel set, leading to birth-densities being higher than death-densities.}. This can be imagined as ``counting'' the number of features in the persistence diagram that are to the left and above of the point on the diagonal with the chosen threshold density. In Fig.~\ref{fig:3_sample_persistence} we illustrate this with three examples, at threshold densities of 0.2, 0.5 and 0.7, leading to respective Betti numbers of 5, 35 and 14.

Fig.~\ref{fig:3_sample_persistence} also illustrates the concept of \emph{persistence} and topological noise. Persistence refers to the stability and lifetime of a feature. It is simply the difference between the birth- and the death-density, and thus quantifies a density range at which it exists in the field. Features with high persistence are long-lived, stable and prominent (e.g. an isolated high-density island), whereas a low persistence value indicates features that are short-lived or transient, and can sometimes be mere noise. In Fig.~\ref{fig:3_sample_persistence}, the blue shaded region close to the diagonal indicates this topological noise, and the red shaded region in the upper part of the diagram marks several points of high persistence. In particular the high-persistence points are of great relevance for this study, as they trace the most prominent features of the cosmic web (clusters, filaments, and voids).

The persistence calculation is done on the basis of decreasing superlevel sets of the DTFE density contrast (equation~\ref{eq:delta}). Essentially, the simplices of the Delaunay triangulation are sorted according to their density value. The nested hierarchy of superlevel sets of the density field are generated by gradually decreasing the density threshold. The homology of these nested superlevel sets is calculated using the Persistent Homology Algorithms Toolbox (PHAT) by~\citet{bauer2014clear,bauer2017phat}. PHAT version 1.2.1 is used for all calculations in the present study. It returns a list of independent features with associated dimension, birth-density and death-density. In order to facilitate the homology computation by the PHAT toolbox, we use a grid that is a slightly dithered version of a completely regular grid, with slightly perturbed positions of the completely regular grid~\citep[see][]{bendich2010computing}, this avoids degenerate point constellations and the resulting non-unique structures.

\subsection{Persistence visualisation}\label{sec:persistencevisualisation}
Depictions of persistence diagrams include one more simplification: instead of plotting the persistence points as points, we provide a persistence histogram, showing the density of points per $h^3$Mpc$^{-3}$ at a certain birth/death density. Due to the large number of persistence points (more than 200000), depicting them as points is problematic, as separate points would be impossible to discern, hence the move to indicate the density of points instead. Due to the wide range of birth/death densities over which structures are present at this stage, this wide range is also present in the persistence diagram. With the hierarchical process of structure formation, there is also a very large number of small-scale structures, as opposed to much fewer large-scale (persistent) structures. In the persistence diagram this topological noise occurs with many more persistence points being located close to the diagonal (where birth- and death density are similar) than further away (in the region of high persistence). The orders of magnitude difference makes logarithmic scales both in the axes and the colour bar necessary. This behaviour, as well as the roughly triangular shape, is similar in all three dimensions.

%% file: 3_results1.tex
\section[Homology of the cosmic web in LCDM cosmology: z = 0]{Homology of the cosmic web in \\ $\bm\Lambda$CDM cosmology: $\bm{\MakeLowercase{z} = 0}$}\label{sec:resultsz0}
The topology of the \lcdm cosmic web at $z=0$ is used as base reference for the other snapshots. We first discuss the overall topology of the \lcdm mass distribution in terms of the one-dimensional Betti curves $\beta_i$ (at superlevel density threshold $\log(\delta+1)$). Subsequently, we turn towards the persistence diagrams for a detailed investigation of the multiscale structure and connectivity of the cosmic web. It allows us to identify and establish the relationship between the physics of the structure formation process and the topological characteristics of the cosmic web.

\begin{figure*}
 \centering
 \includegraphics[width=0.90\textwidth]{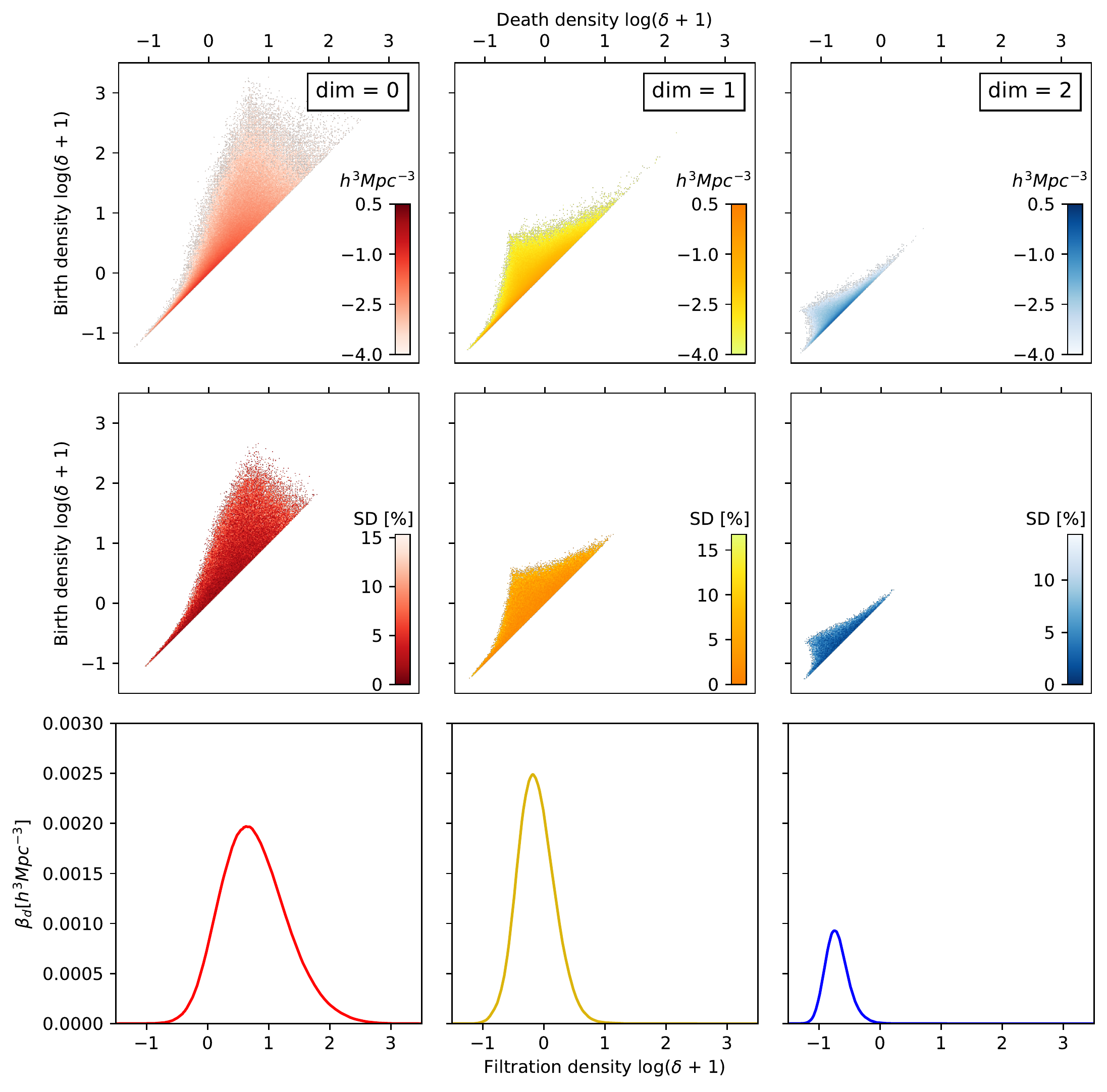}
 \caption{\textbf{Redshift $\bm{z=0}$ persistence and Betti curves.} In the top row from left to right, we depict the logarithmic persistence point density for topological features of dimensions zero, one and two (in red, yellow and blue) for one of the five independent runs. The birth- and death-densities on the axes also follow a logarithmic scale. he centre row shows the standard deviation (SD) of the persistence diagrams of the five runs (note that darker shading indicates higher agreement, i.e., lower SD). The bottom row shows the corresponding Betti curves, also with logarithmic density scale.
 }\label{fig:4_persistence_betti_z7}
\end{figure*}

\subsection{Betti curves: global homology of the cosmic web}
In the third row of of Fig.~\ref{fig:4_persistence_betti_z7} we present the redshift $z=0$ Betti curves for dimensions zero, one and two (left to right). The three panels share the same density axis to facilitate comparison between the different Betti curves. For all three topological elements, -- islands (dimension zero), loops of filaments/tunnels (dimension one) and voids (dimension two) -- we find a comparable behaviour. For all dimensions, the Betti curves are peaked functions centred around a maximum, indicating the density at which the (superlevel) density field contains the highest number of independent topological features components. The Betti curves fall off towards zero towards both lower and higher density thresholds\footnote{The zero-dimensional Betti curves actually fall of to one, resulting in one connected component after the lowest threshold is reached, with a theoretical death-density of $-\infty$. As this is always the case (regardless of redshift) and to allow the presentation of persistence diagrams on logarithmic scales, we ignore this single point. The behaviour of Betti curves at the lowest thresholds becomes more relevant when treating observational data with non-periodic behaviour. Research in this direction is currently being finished and prepared for publication~\citep{wilding2021b}.}. The decrease at the high density wing indicates that the corresponding features become increasingly rare towards higher density levels. As we proceed to even lower density levels, different components start to merge into ever larger agglomerates. Ultimately all components merge into one percolating structure, and all individual features disappear entirely.

While the Betti curves display the same generic behaviour, the density ranges differ considerably. The two-dimensional void population reaches significance only at density levels below the average density, $\delta=0$. By contrast, a distinct presence of zero-dimensional islands is seen to characterize the density field over more than two orders of magnitude: we find islands at $\delta \approx 100-1000$, whereas their numbers are skewed strongly towards higher density levels. Nonetheless, we even find some at $\delta \approx -0.5$. The highest number of individual objects is that of the tunnels and filaments. They dominate the density field around the average density, with a slight skew towards lower density levels. 

Fig.~\ref{fig:5_betti_residuals} has superimposed the curves $\beta_0$ (red, dash-dotted), $\beta_1$ (yellow, solid) an $\beta_2$ (blue, dashed) in order to better appreciate the systematic differences between the Betti curves. The overlap ranges of the different curves provide substantial information on the formation process that has produced the density field. For a more detailed discussion, it is helpful to infer quantitative information on the Betti curves. Towards this end, we parametrize the curves by a skew normal distribution.

\subsubsection{Betti curve parametrization}
\label{sec:betticurveparam}
The Betti curves in Fig.~\ref{fig:4_persistence_betti_z7} appear to be largely symmetric in terms of the logarithm of the density field contrast $\delta$, be it with some moderate level of skewness~\citep{pranav2015thesis,pranav2021topology}. The moderate skewness in terms of the logarithmic density contrast is related to the overall near log-normal density distribution of the evolved nonlinear cosmic mass distribution~\citep{colesjones1991}. Within this context, the skewness of each of the Betti curves can be understood from the realisation that it is the evolved manifestation of the symmetric log-normal density distribution for each of the structural components (i.e. of the matter concentrations, filaments and voids). Based on this observation, we use the first-order term of the normal distribution expansion, the skew normal distribution~\citep{ohagan1976bayes,azzalini1985class}. The function is the product of the standard normal distribution function $\phi$ and its cumulative distribution function $\Phi$. 

\begin{equation}\label{eq:skewnormal}
    f(x | \xi, \omega, \alpha, c) = \frac{2c}{\omega}\phi\left( \frac{x - \xi}{\omega}\right)\Phi\left(\alpha\frac{x - \xi}{\omega} \right)\,.
\end{equation}
In this expression, $x$ is the filtration density $\delta$. $\xi$ is the location parameter and $\omega$ the scale factor of the distribution, while $c$ is a normalisation constant. The value of $\alpha$ parametrizes the shape of the curve and relates directly to the skewness of the distribution: when $\alpha > 0$ the curve is right skewed, when $\alpha < 0$ it is left skewed. For the fitting routine (\texttt{scipy.optimize.curve\_fit}), we use Python's implementation of the normal distribution from \texttt{scipy.stats}. The uncertainty in the value of the parameters is estimated from the variation between the five different realisations.

\begin{figure}
\includegraphics[width=\columnwidth]{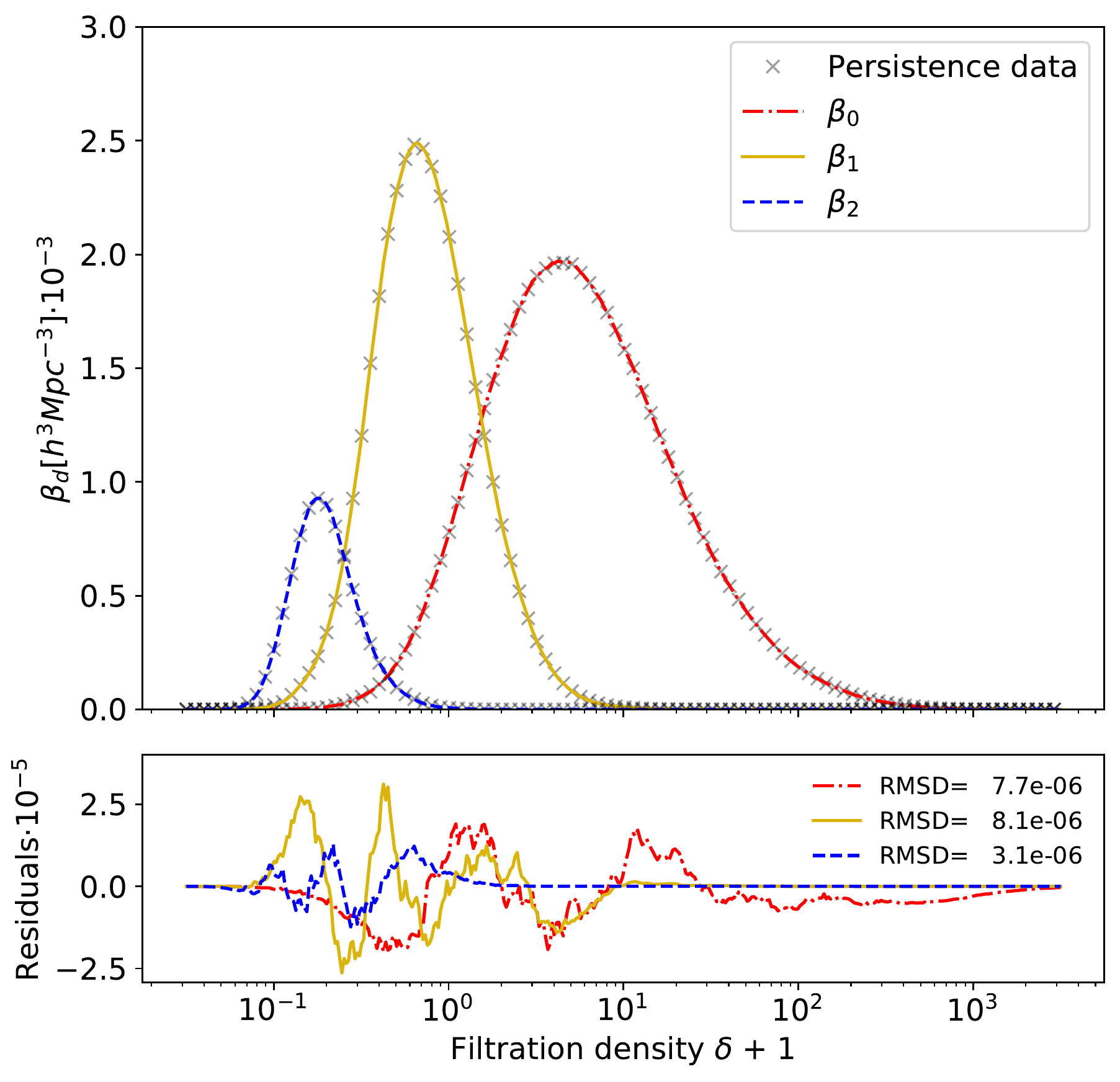}
\caption{\textbf{Betti curve comparison at redshift $\bm{z = 0}$.} The Betti curves of dimension zero, one and two (red, yellow and blue lines), in logarithmic scale (top), superimposed. The crosses indicate the Betti number at specific filtration densities that where used as a basis for the parameter estimation (Table~\ref{tab:peak_densities}). Bottom panel: Residuals of the Betti curve skew normal fits, with the root-mean-square deviation (RMSD) given in the legend.}
\label{fig:5_betti_residuals}
\end{figure}

Following the above, we fit a skew normal distribution to the Betti curves, yielding four characteristic parameters. The fit works very well, as evidenced by the residuals in the bottom panel of Fig.~\ref{fig:5_betti_residuals} and the low root-mean-square deviation.

\begin{table*}
	\centering
	\caption{\textbf{Betti peak positions in $\Lambda$CDM simulations.} The filtration densities are based on $\delta + 1$ for redshifts $z = 0$, $1$ and $3.8$ in columns two to four from the left, and compared to a Gaussian random field based on $\nu = \frac{\delta}{\sigma}$ (transformed to mean-zero densities, see equation~\ref{eq:f-sigma}) after smoothing on a scale of \unit[2]{\hMpc{}}. This is discussed in more detail later in Section~\ref{sec:gaussian}. The uncertainties are the combined uncertainties from the five runs, and calculated from the bin size of the Betti curves.}
	\label{tab:peak_densities}
	\begin{tabular}{llllc  ll}
	    \hline
	    Dim & \multicolumn{3}{c}{$\delta+1$}   &                 &     \multicolumn{2}{c}{$\frac{\delta}{\sigma}$}  \\
		\hline
		    &    $z = 0$   &  $z = 1$     &   $z = 3.8$     & ~~~~~~  &   \lcdm   &    GRF    \\
		\hline
 0 &  4.4$\pm$0.2  &  4.1$\pm$0.2  &  1.90$\pm$0.05  & ~~~~~~  &  2.0$\pm$0.1  &  $\sqrt{3} \approx 1.7321$    \\
 1 &  0.64$\pm$0.02  &  0.85$\pm$0.02  &  0.93$\pm$0.03  &  ~~~~~~ &  -0.10$\pm$0.03  &  0   \\
 2 &  0.181$\pm$0.005  &  0.297$\pm$0.009  &  0.53$\pm$0.02  &  ~~~~~~ &  -1.23$\pm$0.02  &  $-\sqrt{3} \approx -1.7321$   \\
		\hline
	\end{tabular}
\end{table*}

Among the parameters of the skew normal distribution, the mean $\mu$ and standard deviation $\sigma$ can be inferred directly from the identities

\begin{align}\label{eq:skewmean}
  \mu & = \xi + \omega\mu_z\,, \quad \text{and} \\ 
  \ \nonumber\\
\sigma & = \sqrt{ \omega^2 \left( 1 - \frac{2\delta^2}{\pi} \right) }\,,
\end{align}
\noindent in which
\begin{equation}
\mu_z =  \sqrt{\frac{2}{\pi}\delta}  \,,\qquad \delta = \frac{\alpha}{\sqrt{1 + \alpha^2}}\,.
\end{equation}

\subsubsection{Betti curves and structural connectivity}
A few observations with respect to the Betti curves in Fig.~\ref{fig:5_betti_residuals} bear directly on the connectivity of the various topological features.

The first observation is that there are clearly distinguishable density regimes over which the topology is almost exclusively dominated by only one of the topological features. The regions have substantial overlap, in which we can distinguish two or more different topological features in the superlevel density field. The most substantial overlap regimes concern those between the islands and filamentary loops, i.e., between $\beta_0$ and $\beta_1$, and those between the filaments and voids, i.e., between $\beta_1$ and $\beta_2$. There is a narrow range, around the average density $\delta=0$, at which we see a significant presence of all three feature classes. 

There is a large density range over which the topology is almost exclusively dominated by zero-dimensional features, i.e., by density islands. The mass distribution for $\delta \gtrsim 5$ is mainly that of disconnected island clusters. It is interesting to note that this is approximately the density contrast corresponding to density enhancements undergoing gravitational contraction~\citep{gunn1972infall}. On the low density side we find a similar behaviour with respect to the $\beta_2$ curve characterizing the presence of voids: below $\delta \approx -0.8$ voids are the sole topological features in the density field. Also this we may relate to the dynamics of the structure formation process: voids in the cosmic mass distribution mature and stand out as individual low density basins as they have decreased their density to $\delta \approx -0.8$~\citep{blumenthal1992,shethwey2004}.

For the complex geometry and topology of the cosmic web, the most interesting regime is that where we see a substantial overlap between the Betti curves, most prominently between $\beta_0$ and $\beta_1$. Starting from the maximum of $\beta_0$ at $\delta=3.4$ (see Table~\ref{tab:peak_densities}), which is indeed close to the theoretically expected value of $\delta \gtrsim 5$ for gravitational contraction (and only lower because noisy features at lower densities are considered as well), we see that the number of individual islands/clusters rapidly decreases towards lower density thresholds, while at the meantime noticing from the $\beta_2$ curve a quick rise in the number of tunnels/filaments.

The latter reflects the fact that while individual island clusters merge into ever larger agglomerations, the number of tunnels and filaments connecting these is increasing at an even higher rate. It is the topological signature of the emergence of structure from perturbations~\citep{doroshkevich1970}, in particular that of the cosmic web~\citep{zeldovich1970,bond1996filaments,weygaert2008clusters}: high density ridges get connected into an increasingly percolating structure characterized by filamentary bridges. Ultimately, at the near universal density $\delta=0$, all islands are connected into a single percolating and volume pervading structure, the \textit{cosmic web}.

The overlap between the $\beta_2$ and $\beta_1$ curves differs slightly from that between islands and filaments, in the sense that the corresponding features co-exist over a larger density range (from the perspective of the voids). Physically, it entails the transition from a situation in which the superlevel set at higher density thresholds consists mostly of filamentary bridges to one in which these filaments are absorbed into slabs that fill in the boundaries of underdense void basins. We notice there still is a substantial number of filamentary loops while the superlevel set has attained a near maximum number of fully enclosed voids. Only towards the voids with the lowest densities, we see a rapid decrease of filamentary loops as they get absorbed into their boundary shells.

\subsection[Persistence analysis: multiscale structure and connections in the LCDM cosmic web]{Persistence analysis: multiscale structure and connections in the $\bm{\Lambda}$CDM cosmic web} While the Betti curves provide information on the global topological structure of a density field, insight into the detailed multiscale structure, and the corresponding hierarchical evolution of the field, can only be obtained from the far richer information content of the persistence diagrams. 

The persistence diagrams in Fig.~\ref{fig:4_persistence_betti_z7} reveal the multiscale nature of topological features of various dimensions. We show them in the top row, together with the standard deviation for each bin of the persistence diagrams of the five independent runs (centre row). The points (associated with pairs of birth-death densities) in all three diagrams display a characteristics triangular shaped morphology. They have a firm and broad diagonal base, at which we find the vast majority of detected points, which correspond to low-significance short-lived features. The more interesting region of the diagrams concerns the triangular region. In all dimensions, the hierarchical process of structure formation leads to the convergence of the (birth and death) density ranges towards (for the respective dimension) characteristic values, producing the distinct triangular shape. Typically, it is bounded by the diagonal and two concave edges, with the latter meeting at a sharply defined apex. The (birth,death) pair density along the diagonal is up to four orders of magnitude higher than in the interior of the triangular region. The diagonal points represent topological noise, noisy features that are annihilated shortly after they are born. The better agreement (indicated by the lower standard deviation) for the regions closer to the diagonals is largely due to the high number of persistence points located there. While the standard deviation increases towards the more relevant apex, it is still moderate, although shot-noise starts to appear in regions with exceptionally few persistence points. 

Reflecting the behaviour of the Betti curves, there is a substantial difference in the density range over which the zero-, one- and two-dimensional features -- in the triangular shaped region of significant features -- are found in the persistence diagrams. High density islands expand a density range of more than two orders of magnitude, while filaments and tunnels are found in a much narrower density range of slightly more than one order of magnitude around $\delta \approx 1$. Voids, the two-dimensional features, are mostly confined to an even narrower density range of less than one order of magnitude near $\delta \approx -0.8$.

The interior and concave boundaries of the triangular regions in the persistence diagrams contain a wealth of information on the structure and topology of the corresponding features. This concerns both the overall global distribution of these features, as well as the detailed multiscale structure emanating from the hierarchical evolution of the dark matter distribution. For all three persistence diagrams, we find that one concave boundary tends to have a sharper outline, while the other is more curved and tends to have a more fuzzy and slowly fading outline. Apart from this similarity, we observe telling differences between the diagrams that reflect interesting differences in the multiscale nature and connectivity of peaks and islands, tunnels and filaments and voids. One such difference is visible in the zero-dimensional diagram, where the left-hand wing appears concave at the lowest densities while, separated by an inflection point, exhibiting an almost convex behaviour when approaching the apex. In general, these differences reflect the different density ranges over which the corresponding structural features are born, exist, and die -- global information that is also found in the corresponding Betti curves. In addition, the differences in shape and morphology of the persistence diagrams reflect more profound differences in the multiscale structure and hierarchical evolution of the structural components of the cosmic web.

In terms of their morphology, the most outstanding aspect of the diagrams is the presence of an apex. The existence of such distinct, discontinuous features suggests the presence of a sharp ``phase'' transition in the multiscale embedding of topological features. Also, we find that such a transition works out differently for islands, tunnels and voids.

\subsubsection{Cosmic web formation: island \& filament persistence}
In the case of the zero-dimensional islands (Fig.~\ref{fig:4_persistence_betti_z7}, top-left panel), the apex of the persistence diagram marks the location of the features with the most extreme birth density. They are the islands that have gravitationally formed in and around the highest density peaks in the initial Gaussian field of density fluctuations and which evolved into prominent high-density clusters. These objects reflect the steep Gaussian tail of density peaks~\citep[see][]{bardeen1986,adler1981}. The fact that they are found at such a narrowly defined apex suggests they all disappear at almost the same death density. It is as if these islands get joined -- along with a large number of entities created at more moderate density levels -- into a large agglomerate (or agglomerates) at one particular critical density, $\delta \approx 5$. Interestingly, this is around the density value where matter enhancements decouple from the Hubble expansion and gravitational contraction sets in~\citep{gunn1972infall}.

Turning to the corresponding one-dimensional persistence diagram, we gain more insight into the fate of the disappearing islands in the zero-dimensional diagram. Here we observe that the triangle containing most features has an almost horizontal fuzzy edge. It suggests that there are not many filaments and tunnels that are born above $\delta \gtrsim 5.0$, almost at the same level where we find the apex in the one-dimensional diagram.

While the sharp transition marked by the persistence apexes represents the principal process of cosmic web formation, it may not be surprising that the process is marked by a more varied and richer evolutionary history. We also recognise the imprint of these in the zero-dimensional persistence diagram (Fig.~\ref{fig:4_persistence_betti_z7}). We see that on both sides of the apex, the zero-dimensional persistence diagrams widens. On the low density side, we find a substantial fraction of density islands that merge and disappear at a lower density than that marking the emergence of the cosmic web at $\delta \approx 5.0$. Individual density islands remain in existence even while the major share of mass resides in the cosmic web, to get absorbed into the overall weblike network at a lower density. At the high density side, the persistence diagram is marked by a fuzzy edge. This marks objects that get absorbed by surrounding agglomerations relatively fast after their birth, before these got incorporated in the cosmic web. 

The observed transitions in the zero- and one-dimensional persistence diagrams represent a telling illustration of the birth of the cosmic web. With $\delta \approx 5.0$ marking the level where we notice a characteristic transition in which islands get connected into percolating mass agglomerations, we also observe the birth of many filaments and tunnels. It suggests that the assembly of the merging islands proceeds via the establishment of filamentary connections, along with corresponding tunnels. The zero-dimensional persistence diagram apex indicates that the density concentrations that join into the percolating network of the cosmic web are the ones that decouple from the Hubble expansion and undergo gravitational contraction.  

\begin{figure*}
 \centering
 \includegraphics[width=\textwidth]{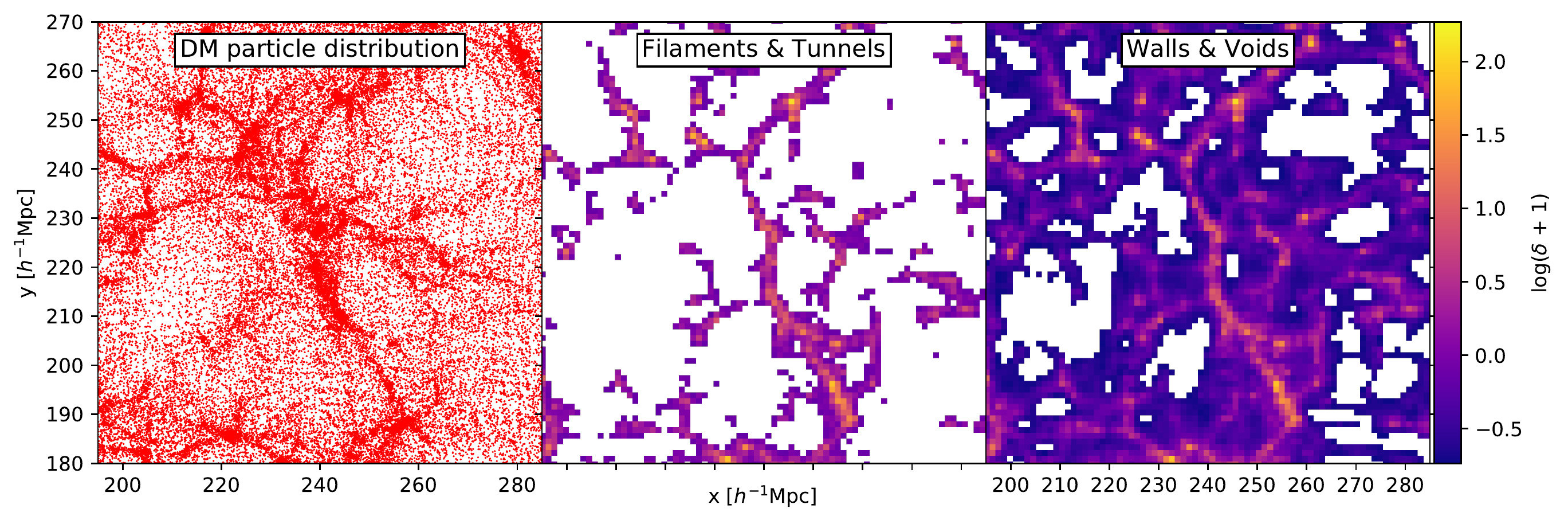}
 \caption{\textbf{The filamentary cosmic web -- zoom.} We show an enlarged region of the density field from Fig.~\ref{fig:1_gadgetdtfe}. To highlight the connecting structure, we compare a slice through the DM particle distribution (left-hand panel) with slices through the density field of the filamentary structure (centre panel) and the wall structure (right-hand panel).
 }\label{fig:6_cosmicweb_zoom}
\end{figure*}

\subsubsection{Void hierarchy: two-dimensional persistence and the void population}
On the low-density side of the matter field, we turn towards the two-dimensional persistence diagram. Its shape differs to that of the zero- and one-dimensional diagrams. It has a sharp apex that marks a narrow ridge of void birth densities around $\delta \approx -0.8$. This is indeed the characteristic density for voids in the galaxy and matter distribution~\citep[see e.g.][]{blumenthal1992,shethwey2004,weygaertplaten2011,weygaert2016voids}. Comparison between the zero- and two-dimensional diagrams therefore reveals that whereas cluster peaks and conglomerates possess a high diversity of densities, voids tend to have a largely similar underdensity. 

A particularly outstanding aspect of the two-dimensional diagram is the sharp apex. It delineates an indentation towards lower birth density levels. It is a reflection of the fact that individual deep voids are non-existent. More towards the right, we encounter voids at such low densities. They tend to be the deepest pits in a larger void complex of a more moderate average density. Evidently, soon after they appear as individual topological features they disappear as they fill up with decreasing density threshold. Also some shallower voids can be discerned, emerging at density levels $\delta > -0.7$. However, these tend to be substantially closer to the diagram's diagonal. Most of these are small shallow void regions near the boundary of large void regions~\citep[see e.g.][]{shethwey2004,hidding2016}.

In summary, the above reveals that at any one cosmic epoch, most significant -- topologically identified -- voids are the ones that show up at a density threshold $\delta \approx -0.8$. At a higher density level, most of these individual voids are embedded and connected in a larger underdense depression, a percolating region that grows in extent towards the higher density levels that demarcate these regions.

It is highly interesting to realise that the two-dimensional multiscale topological structure that we just described is a reflection of the known hierarchical evolution of the void population. The characteristic density ridge in the persistence diagram at $\delta \approx -0.8$ is a reflection of the fact that voids become truly non-linear as they undergo shell crossing, i.e., when their interior mass elements overtake the outer layers~\citep{blumenthal1992,shethwey2004} \footnote{Ideally, $\delta=-0.8$ is the non-linear density of isolated spherically symmetric voids, $\delta_v=-2.81$ the corresponding linear extrapolated underdensity for a shell crossing void~\citep[see][]{shethwey2004}}.~\cite{blumenthal1992} pointed out that it is these matured voids that are the ones found in the matter and galaxy distribution, which~\cite{shethwey2004} translated into a theory for the hierarchically evolving void population~\citep[see also][]{dubinski1993}.

\begin{figure*}
 \centering
 \includegraphics[width=\textwidth]{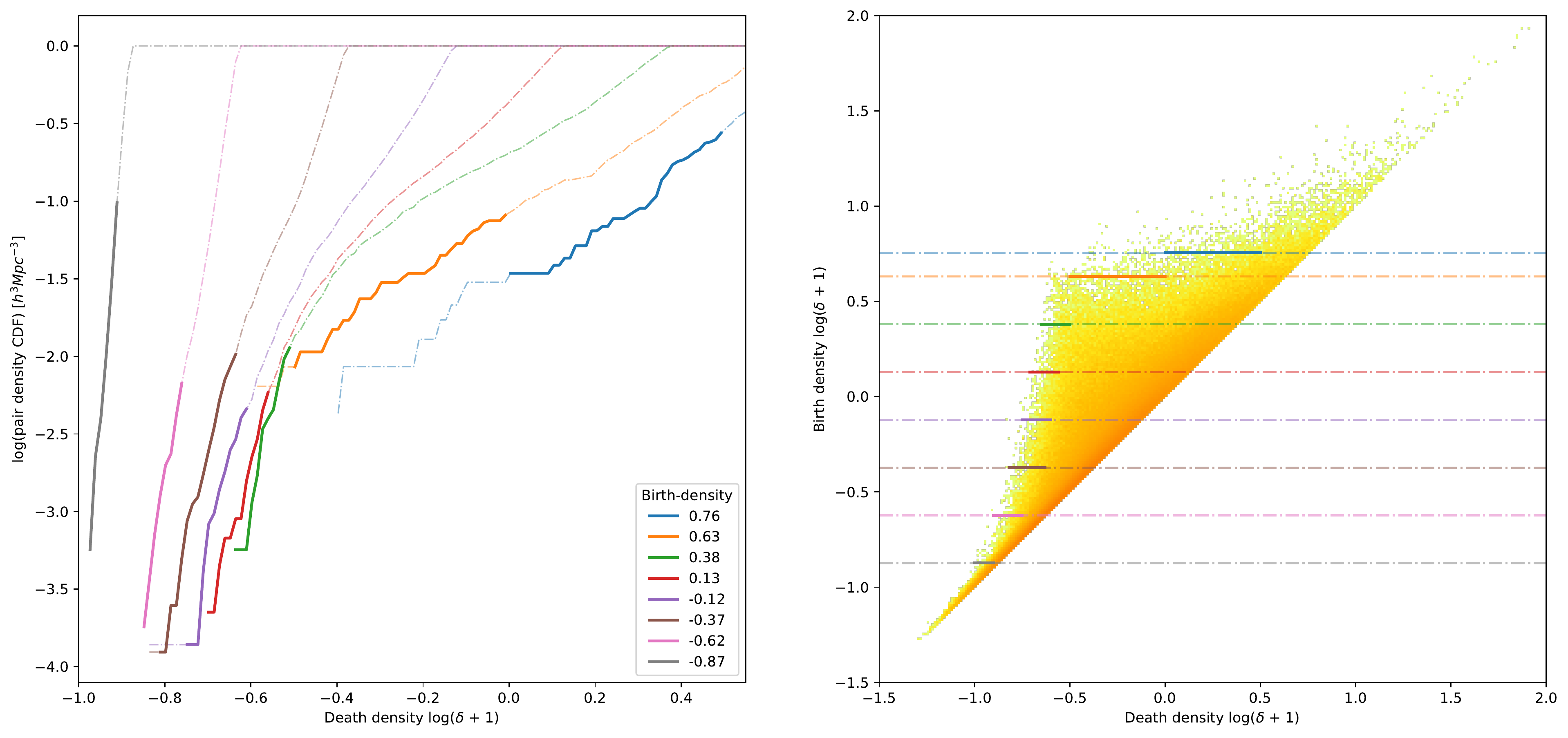}
 \caption{\textbf{Redshift $\bm{z=0}$ marginal death CDF.} We show the marginal cumulative distribution function (CDF) of persistence points for a set of constant birth densities and for increasing death densities (left panel, the $\log(\delta+1)$ birth densities are indicated in the legend). The right panel shows the corresponding persistence diagram, with the birth densities of the CDFs indicated in the same colour as in the left panel. Highlighted (with lines in bold in both panels) are the regions where the CDFs intersect the boundary of the persistence diagram, which is mirrored in a steep rise of the distribution functions, after which they level off. This levelling-off is perceptibly stronger for the curves at high birth densities.}\label{fig:7_marginpersistence}
\end{figure*}
\subsubsection{Filaments and tunnels: the one-dimensional persistence diagram}
Armed with the insight provided by the zero-dimensional diagram on islands and the two-dimensional one on voids, we are equipped to establish the relation with the role of filaments and tunnels in the overall mass distribution. These exist at intermediate densities, where the one-dimensional persistence diagram traces the one-dimensional filamentary network (Fig.~\ref{fig:4_persistence_betti_z7}, middle column).

The one-dimensional persistence diagram also displays several distinctive features. It has a rather symmetric shape, it also has an apex, although it is a rather broad one at the tip of slightly concave edges and whose location differs substantially from that of the zero- and two-dimensional diagrams. The apex is located at a formation density of $\delta \approx 5$ and elimination density
$\delta \approx -0.7$

For our focus on the cosmic web, we may argue that the upper edge of the one-dimensional diagram, the nearly horizontal fuzzy border that slopes slightly upward, is of central importance. In a sense, it demarcates the formation of the cosmic web in the form of a percolating network. From our discussion of the zero-dimensional diagram we already learned that it coincides with a sharp topological transition. To better investigate this transition, we highlight the connected structure in Fig.~\ref{fig:6_cosmicweb_zoom} by enlarging a region of the density field shown earlier (c.f. Figs.~\ref{fig:1_gadgetdtfe} and \ref{fig:2_dtfe_level}). The comparison of the filamentary structure (centre panel) with the DM particle distribution (left-hand panel) shows that the careful selection of the threshold (see Section~\ref{sec:persistencevisualisation} for the details) allows the enhancing of a particular component of the cosmic web. This also holds for the depiction of cosmic walls (right-hand panel), although the visualisation using slices can suffer due to the fact that walls intersecting the slice would appear similar to (less dense) filaments. The actual filaments in the centre panel are shown at a critical threshold (which depends on the number of filamentary loops), where a large number of prominent filaments have already connected up while forming tunnels. These filaments and tunnels are born in the narrow density range of $\delta \approx 5$ in which individual high-density islands get merged into one pervasive network. The corresponding connections are established via the filamentary bridges that we see emerging at this narrow range of density levels in the one-dimensional diagram.

We also find that the web-like network is quite fragile and transient. As we proceed to lower density thresholds the network starts to fill up and incorporate walls, filling loops of filaments and turning them into sheets. Once these are joined into a shell, an isolated cavity splits off and is born as a fully enclosed void. This process relates to the left-hand edge of the one-dimensional persistence diagram -- it is the transition marking the formation of voids. From the diagram we infer that it also occurs in a comparatively narrow density range, corresponding to the steep, nearly vertical, edge on the left-hand side of the two-dimensional apex. It delineates the narrow boundary -- at a density of $\delta \approx -0.7$ -- below which nearly all filaments and tunnels die. At that level, we are actually dealing with the remaining tenuous tendrils and interstices in underdense regions. They are the last vestiges and representatives of the filamentary bridges and tunnels that mark the connections between the largest mass concentrations in the cosmic web. 

\subsubsection{Persistence and cosmic structure formation}
Persistence diagrams open up a significantly higher and more profound level of information on the structure formation process than possible with the more global summary statistics like Euler characteristic or Betti numbers. They are unique in their ability to uncover the nature of structural transitions, such as the sharp ``phase'' transitions we found and discussed in the previous paragraphs. While some of these relate to known physical effects, others -- such as the sharp connectivity transition producing the cosmic web -- are in need of further investigation. 

As an illustration of furthering the exploration of the information content of persistence diagrams, Fig.~\ref{fig:7_marginpersistence} provides more details on the (birth,death) process of topological features, by focusing on their marginal CDF (cumulative distribution function). The diagram reveals the density levels at which features born at one particular density threshold finally disappear. We obtain this by assessing the distribution along horizontal lines in the persistence diagrams (see right-hand panel of Fig.~\ref{fig:7_marginpersistence}). In all cases, we find a steep rise, coinciding with a density level at which these features enter the left-hand edge of the persistence diagram (left-hand panel, Fig.~\ref{fig:7_marginpersistence}). In the interior of the persistence diagram, there is a near uniform distribution of densities at which features disappear, translating into a near linear increase of the CDF. This situation changes only near the diagonal, as we get to deal with noisy structure.

From the left-hand panel of Fig.~\ref{fig:7_marginpersistence} we also see the systematic shift of death densities as we proceed from high filament and tunnel birth densities to the lowest birth densities: the last vestiges of filaments and tunnels, that go along with the formation of low density basins, are of a different nature than the prominent filamentary bridges and tunnels that are born as the percolating network of the cosmic web established itself at a density level $\delta \approx 5.0$. Turning to the low density side, in the marginal CDF we see that below birth density $\delta =-0.76$, the filaments/tunnels are hardly significant: they disappear almost at the same level as they are born. Thus at the level where we see the formation of individual voids, there are no longer filamentary tendrils bridging along these regions.

%% file: 4_results2.tex
\section[LCDM cosmic web homology: evolution]{$\bm{\Lambda}$CDM cosmic web homology: evolution}\label{sec:resultsevolution}
Following the detailed analysis of the topology of the \lcdm mass distribution at redshift $z=0$, we address the evolution of that topology in terms of the development of the Betti curves and the persistence diagrams.

To assess the evolution of the cosmic web topology in \lcdm, we analyse the \lcdm mass distribution at 8 redshifts, $z = 3.8$, $2.98$, $2.05$, $1.00$, $0.51$, $0.25$, $0.1$ and $0.0$. We use five different simulation runs to obtain estimates of the variance and uncertainty in the resulting mass distribution at each of the redshifts. 

\subsection{Betti curves: evolving global cosmic web homology}
\begin{figure*}
  \centering
  \includegraphics[width=\textwidth]{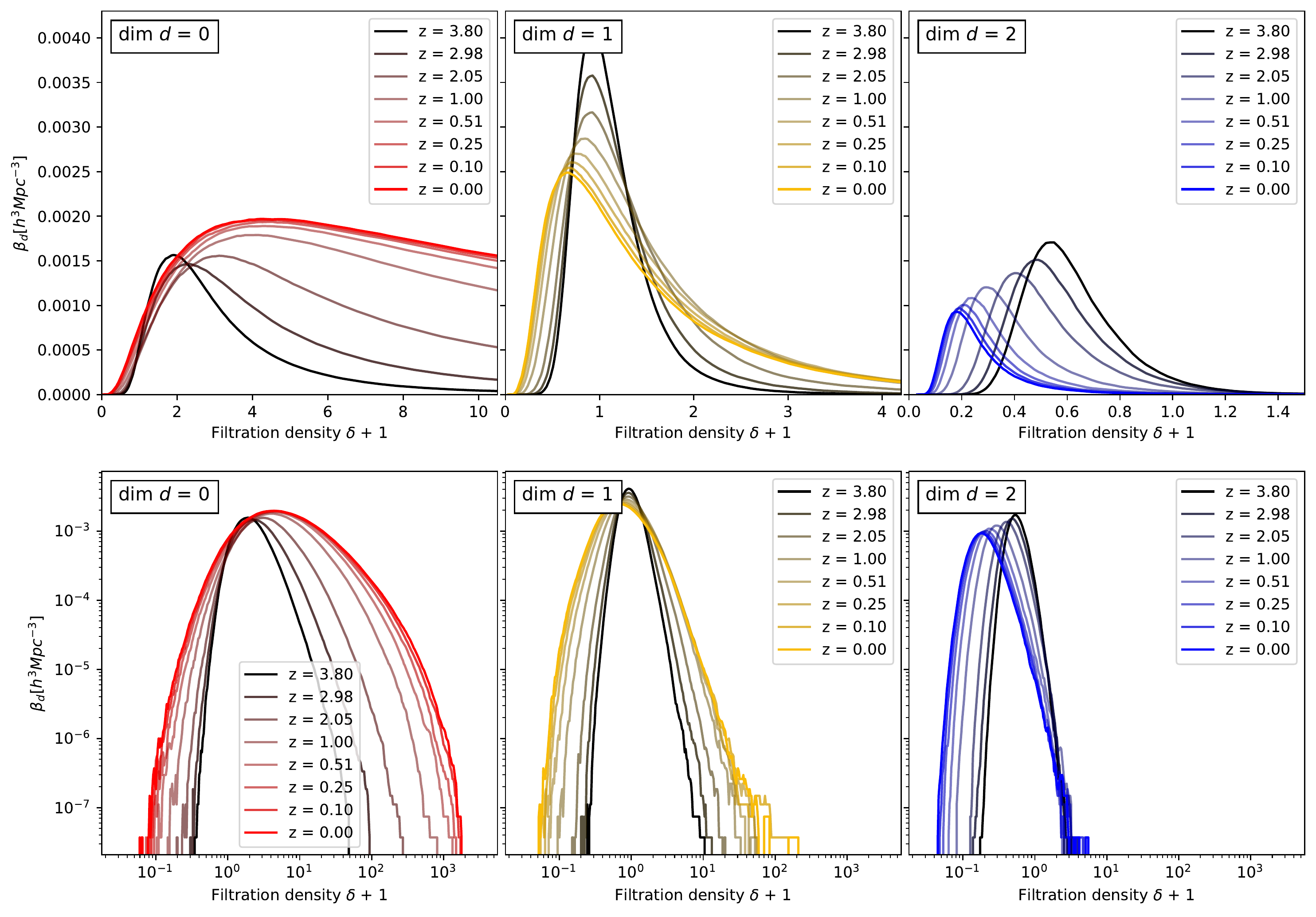}
 \caption{\textbf{Evolution of $\bm{\Lambda}$CDM homology, Betti curves.} Betti curves of dimensions zero, one and two (left to right) for all redshifts, in linear (top row) and logarithmic scale (bottom row). To illustrate the time evolution, the curve corresponding to the earliest snapshot (redshift $z = 3.8$) is the darkest (black) and brightness increases when progressing towards lower redshift, until the curve for redshift $z = 0$ is red, yellow or blue, according to the respective dimension of zero, one and two. We see a clear shift in the position of the curve maxima, as well as in their height.}\label{fig:8_betticurve_evolution}
\end{figure*}

Fig.~\ref{fig:8_betticurve_evolution} presents the evolving Betti curves for the three topological features -- islands (dimension zero), filamentary loops (dimension one) and voids (dimension two). In each of the panels, we superimpose the Betti curves of the corresponding dimension for each of the probed redshifts. The top row panels list the Betti curves with the $x$- and $y$-axis both in linear scale, and the corresponding log-log diagrams are lined up in the bottom row. The evolving topology of the mass distribution can be most straightforwardly appreciated from the log-log plots.
They provide the following observations: 

\begin{itemize}
\item The zero- and one-dimensional Betti curves are systematically broadening as the mass distribution evolves. Both the low density and high density wings are widening, around a maximum that is shifting only relatively weakly. Also the two-dimensional Betti curve is broadening, but only moderately, accompanied by a large systematic shift of the peak towards lower densities.

\item The height of the one- and two-dimensional Betti curves shows a downward trend. By contrast, the zero-dimensional shows an upward trend. 

\item The maximum of all three Betti curves at early times and high redshifts centres around the mean density, i.e., $\delta+1=1$. As the mass distribution evolves, the maximum of all three curves shifts away from the mean density. The maximum of the zero-dimensional curves shifts towards higher densities. The maximum for the one-dimensional curve shifts to slightly lower densities, while the peak of the two-dimensional Betti curve shows a large systematic shift towards lower densities.
\end{itemize}

\subsubsection{Betti curve evolution: quantitative analysis}
To quantify the systematic changes of the Betti curves we assess the evolution of the parameters of the fitting skew normal curves (see equation~\ref{eq:skewnormal}). As discussed in Section~\ref{sec:betticurveparam}, the skew normal curves are fully specified by four parameters, a location $\xi$, scale factor $\omega$ and shape $\alpha$, together with a normalization constant $c$. We determine the values of these four fitting parameters for each of the three Betti curves, at each of the eight analysed snapshots.

\begin{figure*}
 \centering
  \includegraphics[width=\textwidth]{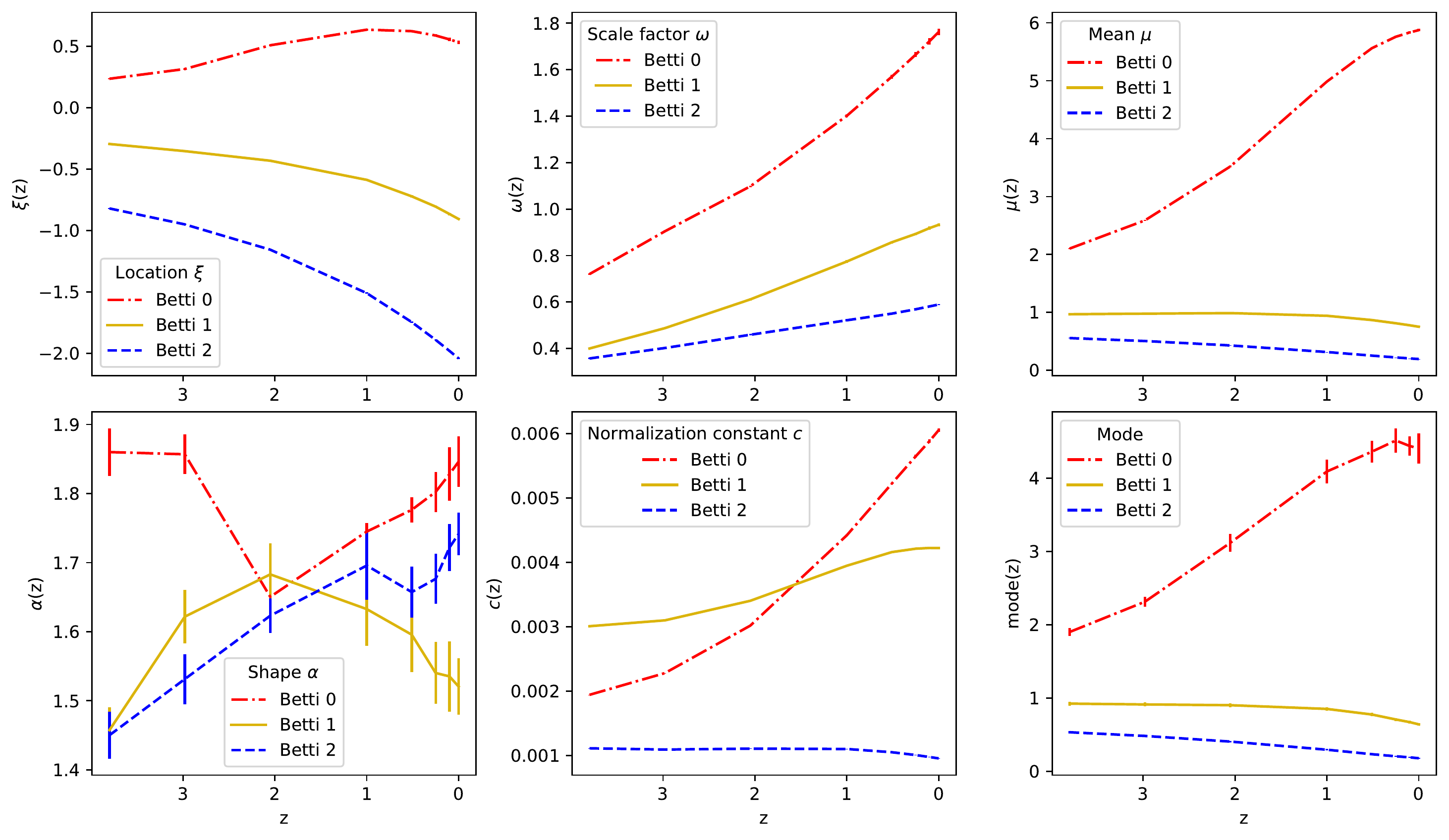}
  \caption{\textbf{Betti curve evolution: parameters.} Fit parameters location $\xi$ (top left), scale $\omega$ (top middle), skew $\alpha$ (bottom left) and scaling constant $c$ (bottom middle) of the Betti curves and their evolution at eight different redshifts, calculated as the mean of five simulation runs. The uncertainties are the combined uncertainties as provided from the fitting routing. Except for the skew $\alpha$, the uncertainties are too small to be visible. Clear trends are apparent, but more meaningful descriptive parameters can be calculated in an additional step and are shown in the right column: the mean $\mu$ (top right) and the mode (bottom right). The mean and the corresponding uncertainty is calculated directly from the fitting parameters. The mode is measured from the Betti curves, with uncertainties as the standard deviation of the five measurements. We point out that uncertainties are always known but only discernible for the zero-dimensional mode).}  \label{fig:9_fit_parameter_evolution}
\end{figure*}

Fig.~\ref{fig:9_fit_parameter_evolution} shows the development of these parameters (left and middle column) as function of redshift $z$. A few systematic
trends immediately stand out:
\begin{itemize}
\item For dimensions one and two, the location parameter $\xi(z) $ (top left-hand) displays a monotonic decrease from high redshift to $z=0$. Over
  nearly the entire redshift range we see an increase for the zero-dimensional location parameter $\xi_0(z)$, nearing a plateau or minor decrease from
  $z=0.5$ to $z=0$.

\item All Betti curves are monotonically broadening. The width $\omega_0(z)$ for dimension zero is most steeply increasing, while the width $\omega_2(z)$ of the two-dimensional Betti curves reveals a moderate growth.

\item The evolution of the shape parameter $\alpha(z)$ is not uniform. For dimension two, we see a monotonic increase of the shape parameter $\alpha_2(z)$. It indicates a continuous increase of the skewness of the void Betti curve towards higher densities, as it shifts from the initial near Gaussian phase towards an ever more stronger non-Gaussian distribution. The shape of the zero- and one-dimensional Betti curves does not reveal major systematic changes, although they both show deviant values at and around $z=2.0$, with a sharp minimum for dimension zero and a maximum for dimension one.

\item Also the normalization constant $c(z)$ reveals characteristic behaviour. While the scaling parameter shows a monotonic and steep increase for the zero-dimensional Betti curve, $c_2(z)$ retains an almost constant value. The Betti curve for loops of filaments reveals a mild increase towards $z=0.0$.
\end{itemize}

\noindent The evolution of one inferred parameter (the mean $\mu_d(z)$) is plotted in Fig.~\ref{fig:9_fit_parameter_evolution} (top right-hand panel). It provides a more direct view of the evolving Betti curves: the development of the mean $\mu_d$ directly reveals the shift of the peak maximum. This is also borne out by the mode\footnote{Which unfortunately cannot be calculated analytically for  the skew normal distribution. Also notice that the uncertainties of the mean are much lower than the uncertainties of the mode. The mean is calculated directly from fitting parameters, whereas the mode is measured from the original curve itself. The uncertainties of the latter depend on the sampling of the curves.}, which we show in the bottom right-hand panel of Fig.~\ref{fig:9_fit_parameter_evolution}. Both panels show a clear increase of the mean and mode of the peak of the zero-dimensional Betti curve, along with a monotonic decrease of that for the two-dimensional Betti curves. The one-dimensional Betti curves indicate a filamentary network that appears to evolve more strongly after $z=1.5$, from which epoch onward we notice a gradual decrease of its characteristic density. 

\subsection[Initial conditions and cosmic web homology]{The $\bm{\Lambda}$CDM cosmic web and Gaussian initial random field}\label{sec:gaussian}
The process of structure formation in the universe proceeds along distinctly different regimes of dynamical development. It starts with the initial field of Gaussian random density and velocity fluctuations. Subsequently, structure evolves from a long linear evolution phase in which it retains a near perfect Gaussian character. The first vestiges of complex structure emerge during the quasi-linear phase, ultimately culminating in the development of highly non-linear collapsed structures and objects in the fully non-linear regime.

Given that the cosmic web and non-linear structure are the product of the gravitationally evolved initial Gaussian conditions, it is interesting to investigate in how far it has retained -- topologically -- the memory of the primordial density and velocity field out of which it arose. In several accompanying studies we analysed in detail the structure and topology of Gaussian random fields~\citep{pranav2019topology,feldbrugge2019,pranav2021topology}.

\begin{figure*}
 \centering
 \includegraphics[width=\textwidth]{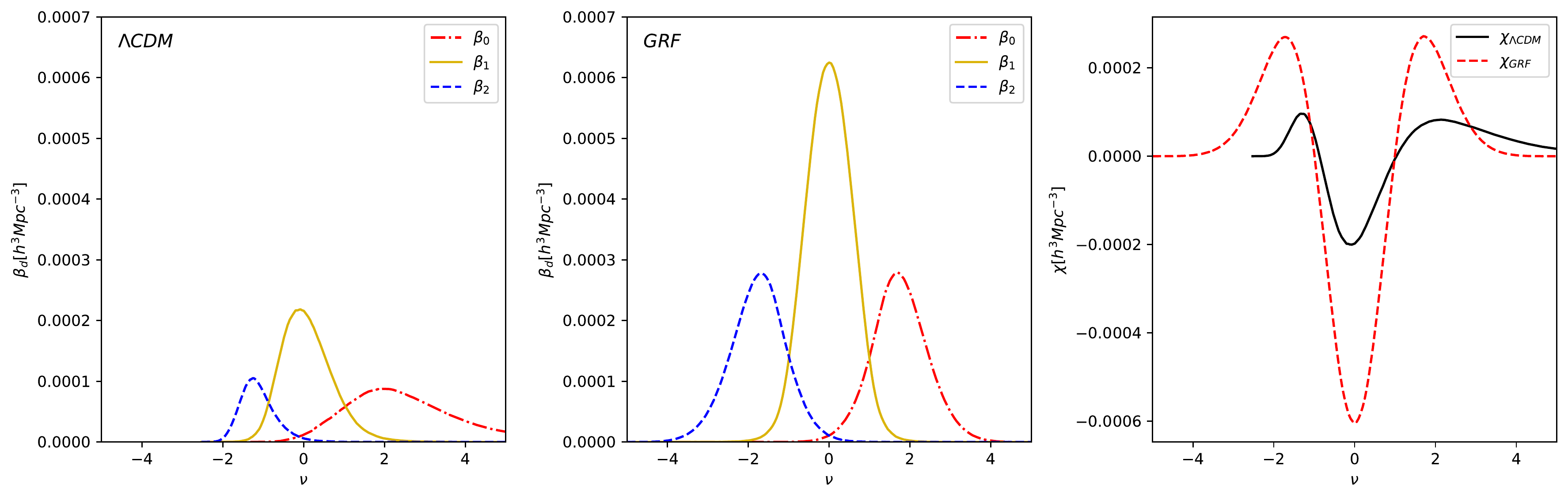}
 \caption{\textbf{Betti curves: $\bm{\Lambda}$CDM vs. Gaussian Random Field.} For dimensions zero, one and two, we show the Betti curves of the \lcdm cosmic web at redshift $z = 3.8$ (averaged over the five runs) on the left and of a Gaussian random field in the middle. In particular the position centred at zero of the one-dimensional Betti curve is similar, as well as the positions of the peaks of the zero- and two-dimensional Betti curves to the right and left, at roughly $\pm\sqrt{3}$. The right-hand panel shows the Euler characteristic for the Gaussian random field and that for the evolved weblike distribution at redshift $z=3.8$ }\label{fig:10_bettieuler_lcdm_gaussian}
\end{figure*}

Fig.~\ref{fig:10_bettieuler_lcdm_gaussian} compares the topology -- in terms of the Betti curves -- of the earliest epoch represented in the simulations at redshift $z=3.8$ with that of a related Gaussian random field. To facilitate comparison, we use the normalized density $\nu$ as filtration quantity,
\begin{equation}
\nu=\frac{\delta}{\sigma}\,.
\end{equation}
To allow a comparison, both fields are smoothed on a scale of \unit[2]{\hMpc{}} using a Gaussian filter. The first observation is that of a prime difference between the symmetric Gaussian field and the non-linear density field. In a Gaussian field the $\beta_0$ and $\beta_2$ curves are mirrored, symmetric images of each other. This reflects the perfect symmetry between underdense and overdense regions in Gaussian fields. Non-linear gravitational evolution evidently breaks the symmetry between underdense and overdense regions. This is clearly reflected in the strong asymmetry between the $\beta_0$ and $\beta_2$ curves in the left-hand panel of Fig.~\ref{fig:10_bettieuler_lcdm_gaussian}.

As a result, the field develops an ever larger asymmetry between underdense and overdense regions. While underdense regions are confined to a density deficit in the limited range of $-1 < \delta < 0$, overdense regions develop a long tail of almost unconstrained overdensities, such as massive clusters of galaxies with overdensities in excess of $\delta \approx 1000$. Gravitational evolution leads to the development of a field with an increasingly non-Gaussian character. In the strongly non-linear situation this can be reasonably approximated as a log-normal field~\citep{colesjones1991}. 

Topologically speaking, we find that at $z=3.8$ the $\beta_0$ and $\beta_2$ curves are strongly deformed, skewed and shifted versions of the corresponding Betti curves in the linear-regime Gaussian field. Whereas the order of the Betti curve maxima remains the same, their exact positions help to illustrate the differences. In the case of the Gaussian random field they are located at $\nu = -\sqrt{3} \approx -1.732$, $0$ and $+\sqrt{3} \approx 1.732$. For the evolved mass distribution at $z=3.8$ we find that the maxima of the $\beta_0$, $\beta_1$ and $\beta_2$ curves have shifted to $\nu 
\approx 2.0$, $-0.1$ and $-1.23$ (see also Table~\ref{tab:peak_densities}). 

The $\beta_0$ curve has developed a long high-density tail reflecting the formation of the gravitationally contracted and collapsed mass concentrations. The $\beta_2$ curve shows that the void population is much smaller than that of the wide spectrum of overdense mass concentrations. Evidently, almost by definition it remains within a narrow density range. As a consequence of the hierarchical evolution of voids -- through merging of smaller voids into ever large ones -- the number density of voids (and hence the area below the Betti curve) is decreasing. In combination with the fact that voids occupy most of the cosmic volume~\citep[see e.g.][]{weygaertplaten2011,cautun2014}, and hence do not leave space for additional ones, the implication is a decrease in the number of voids. 

\begin{figure*}
 \centering
 \includegraphics[width=\textwidth]{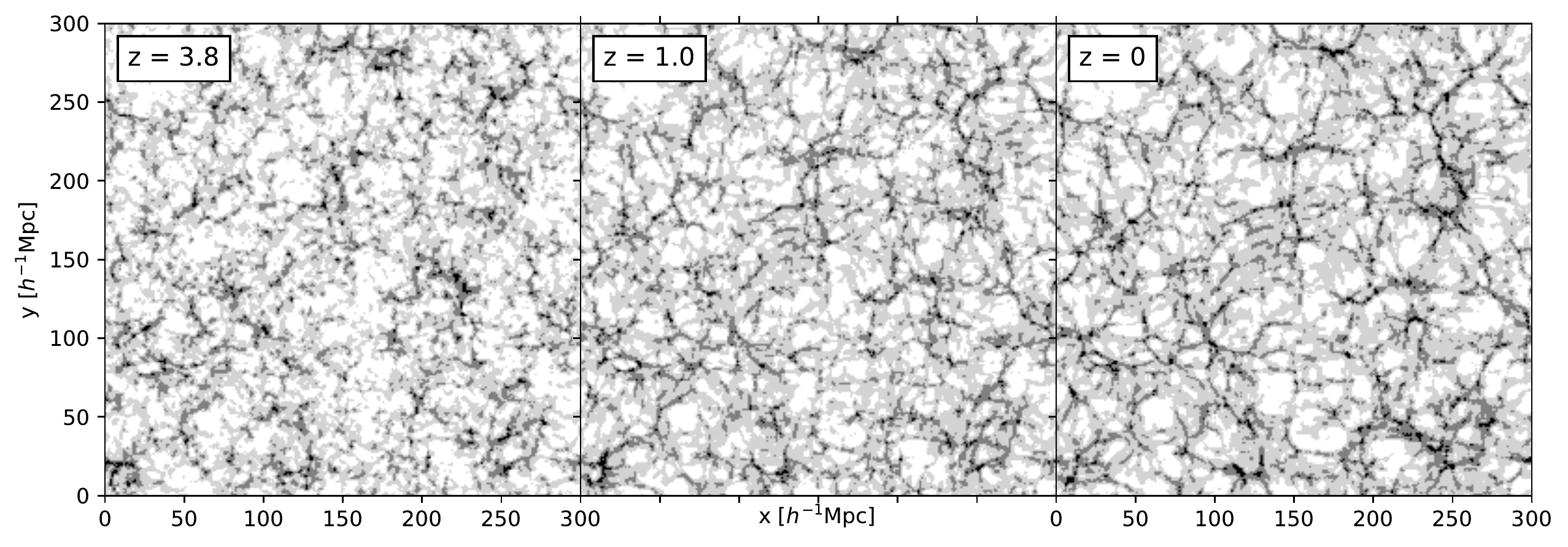}
 \caption{\textbf{Shaded slices through the density field.} We show the evolution of the cosmic web for decreasing redshift from left to right. Structural components are emphasised by shading the pixels depending on the Betti numbers. The cut-off densities are the same as in Fig.~\ref{fig:2_dtfe_level} and depend on the maxima of the Betti curves (Table~\ref{tab:peak_densities}). High density regions are coloured black, regions of intermediate density are coloured dark-grey, low density regions are coloured light-grey and the regions of lowest density are coloured white. We associate black with clusters, dark-grey with filaments, light-grey with walls and white with voids.}\label{fig:11_dtfe_greyscale}
\end{figure*}

The development of the filament and tunnel population as represented by the $\beta_1$ curve appears to entail a more modest evolution. The $\beta_1$ curve at $z=3.8$ still resembles the curve of the Gaussian initial conditions, though now modestly skewed with a slightly longer tail towards higher densities. For lower densities it appears to fall off towards 0 faster than in the initial Gaussian field. This suggests that in particular the population of higher density filamentary bridges, a key element in establishing the cosmic web, is gradually becoming a more prominent aspect in the cosmic mass distribution. This is in line with our view of the dynamical evolution and buildup of the cosmic web~\citep[see e.g.][]{weygaert2008clusters}.

We know that the Betti numbers are intimately related to the Euler characteristic~\citep[see e.g.][]{weygaert2011alpha,pranav2019topology}. The Euler characteristic $\chi$ is the alternating sum of Betti numbers,
\begin{align}\label{eq:euler}
    \chi &= \sum_{k=0}^{\infty} \left( -1\right)^k \beta_k \nonumber \\
        &= \beta_0 - \beta_1 + \beta_2\,.
\end{align}
The right-hand panel of Fig.~\ref{fig:10_bettieuler_lcdm_gaussian} presents the comparison of the Euler characteristic $\chi$ for the Gaussian initial conditions (red, dashed) with the evolved weblike distribution at redshift $z=3.8$ (black, solid). The Euler characteristic at $z=3.8$ has clearly evolved away from the well-known symmetric shape for a Gaussian random field~\citep[see][for the analytical expression]{adler1981,bardeen1986,hamilton1986}\footnote{Strictly speaking, the symmetric expression for the Euler characteristic of Gaussian random fields is only valid for compact manifolds without boundary. The correct expression for any (more realistic) configuration is given by the Gaussian Kinetic Formula~\citep{adler2009random,pranav2019topology}.
}. Instead, we see a narrow low-density wing and a broad high density wing.

\subsection{Topological visualization of density fields}\label{sec:densityvisualisation}\label{sec:bettivisual}
One aspect of Betti curves that we may use to provide an informative topological visualization of the mass distribution is the finding that the characteristic topological features -- voids, loops of filaments and clusters -- typically dominate the mass distribution over specific density ranges. We infer this directly from the fact that the corresponding Betti curves delineate different ranges over which they peak. In other words, Betti curves of different dimensions dominate at characteristic density level, which implies that the mass distribution at different density levels is dominated by different structural components (Figs.~\ref{fig:2_dtfe_level} and~\ref{fig:6_cosmicweb_zoom}).

The indication that each of the specific topological features is dominant over a specific density range suggests the possibility to -- at least roughly -- visualize the occurrence of islands, filaments and tunnels, and voids and walls by identifying typical density thresholds and plotting the corresponding superlevel sets. Fig.~\ref{fig:2_dtfe_level} shows the superlevel sets corresponding to density levels equal to the maxima of the zero- (top row), one- (medium row) and two-dimensional (bottom row) Betti curves. The evolution of the evolving structural elements of the cosmic web can be appreciated from the three panels in each row: the left-hand panel shows the high redshift configuration at $z=3.8$, the middle panel that at a medium redshift $z=1.0$ and the right-hand panel the low redshift situation at the present epoch, $z=0$. The values for the densities at the corresponding Betti curve maxima are listed in Table~\ref{tab:peak_densities}.

The topologically selected patterns elucidate the role and development of clusters and islands, filaments and tunnels, and voids and walls, in defining the cosmic web. The first structures to emerge in the cosmic matter distribution are the peaks and the matter islands forming around them. While they represent rare mass concentrations at high redshift, from $z=1$ onward their distribution reveals  a spatial organization along weblike patterns, where they are found in the most prominent filaments and walls of the cosmic web. Along with this, the accompanying development of the intricate filamentary and wall-like structures reveals the hierarchical buildup of the spine of the cosmic web~\citep[see][]{aragon2010spine,cautun2014}. At the lowest threshold, corresponding to the dominance of voids, we see that the mass distribution is evolving from one with large disconnected weblike patches into one that consists of a percolating foamlike network permeating the entire cosmic volume. At this level, the mass distribution is dominated by walls and voids, defining a landscape that is indented by void cavities. The void population is evolving hierarchically from one of a large number of smallish underdense regions to one of a considerably lower number of much larger void regions~\citep{shethwey2004}. 

\bigskip
Fig.~\ref{fig:11_dtfe_greyscale} shows a variation on the topological segmentation of the mass distribution. It combines the information of the three Betti curves in one image, in which the shade is determined by the dominant Betti number/topological component. It produces a natural segmentation, in which connected high-density regions are represented by black shades, intermediate density regions with the filamentary structure are shaded dark grey, and low density regions corresponding to walls light grey, while the lowest density regions -- the voids -- are shown in white.

\begin{figure*}
 \centering
 \includegraphics[width=\textwidth]{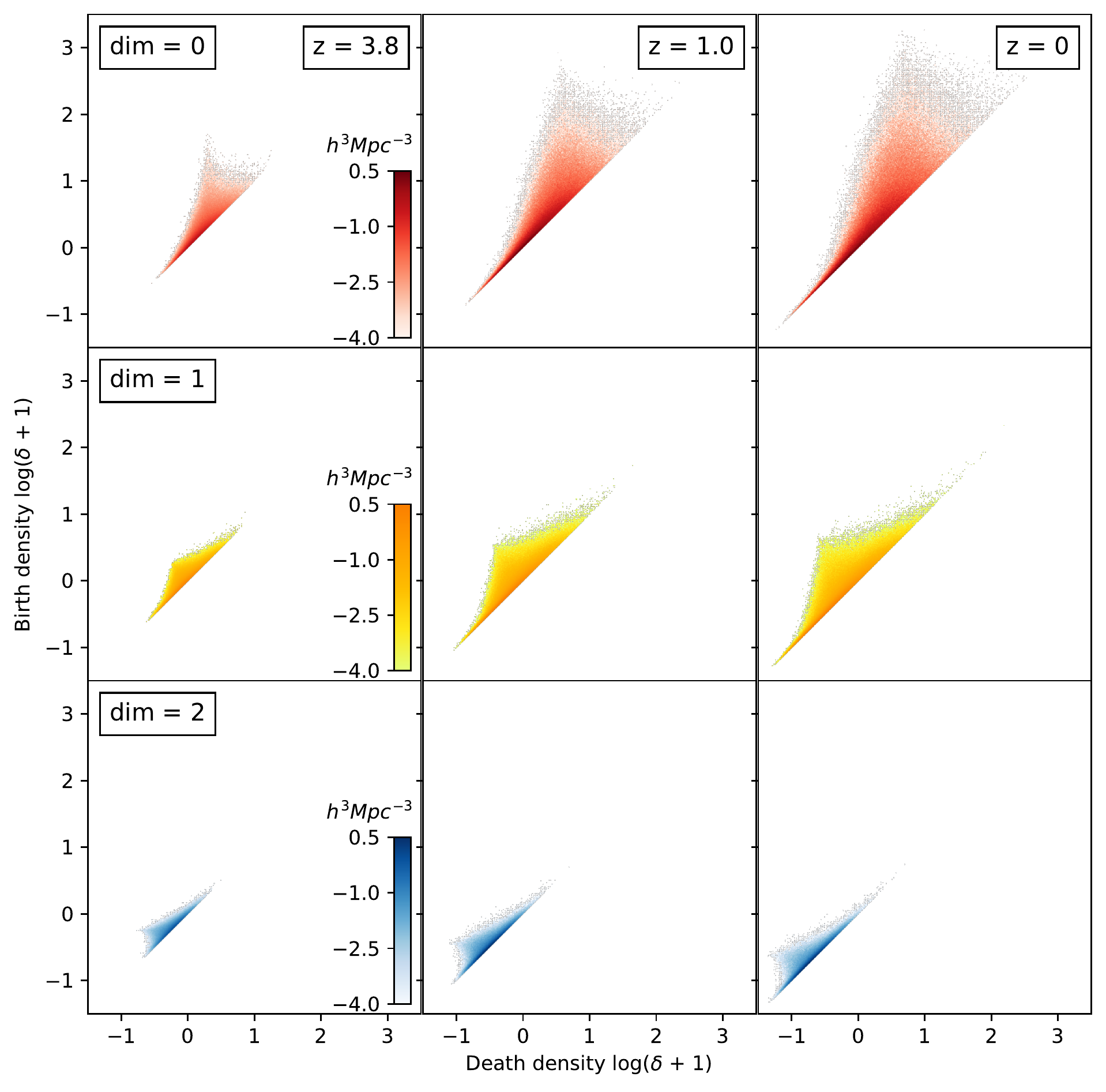}
 \caption{\textbf{Evolution of $\bm{\Lambda}$CDM homology, persistence histograms.} The rows from top to bottom depict the logarithmic persistence point density of dimension zero, one and two (in red, orange and blue). From left to right, the panels have redshift $z = 3.8$, $1$, $0$. The axis limits and scale are the same for all diagrams, to allow easier comparison, even though (particularly for the diagram of dimension two) large parts of the plotting areas do not contain information. We refer to the text concerning what information on the hierarchical formation can be deduced from these diagrams.}\label{fig:12_persistence_evolution}
\end{figure*}

\subsection[Evolution of persistence of the LCDM cosmic web]{Evolution of persistence of the $\bm{\Lambda}$CDM cosmic web}\label{sec:persistence_web}
Persistence diagrams provide detailed information on the evolving multiscale structure of the mass distribution. As such the evolving persistence diagrams form a direct reflection of the intricate hierarchical buildup of structure. 

The typical evolution of the persistence diagrams of the \lcdm mass distribution is shown in Fig.~\ref{fig:12_persistence_evolution}. It shows the zero-, one- and two-dimensional persistence diagrams for three redshifts, the high redshift of $z=3.8$, the medium redshift $z=1.0$, and the present epoch $z=0.0$. A quick first glance at Fig.~\ref{fig:12_persistence_evolution} reveals that: 

\begin{itemize}
\item To first order, the persistence diagrams -- for dimensions zero, one and two -- retain their triangular shape as the cosmic mass distribution evolves. The principal evolutionary trend is a gradual uniform expansion of the triangular core region. This ``expansion'' of the persistence diagrams is a reflection of the gravitationally evolving density field. It leads to the emergence of a growing population of topological features, whose characteristic density spans a continuously increasing range of values. 

\item The uniform expansion of the persistence diagram translates itself in a stretching of the range of birth and death densities of features in the cosmic matter distribution, as well as in their persistence values. 

\item In addition to their widening, we see a shift of the centre of the persistence diagrams' triangular core region. This shift is a clear hallmark of the non-linear hierarchical evolution of the cosmic mass distribution. 

\item In terms of the expansion and shift of the triangular core region of the persistence diagram, there is a marked difference between the zero-, one- and two-dimensional persistence diagrams.

\item The triangular core of the zero-dimensional persistence diagram of islands and matter clumps shows a strong size evolution along with a marked shift. It expands by at least an order of magnitude from $z=3.8$ to $z=0.0$, representing the emergence of islands and clumps whose density contrast $\delta$ is at least 10 to 100 higher than at $z=3.8$. We also observe an increasingly skewed morphology with a centre that shows a strong and systematic shift away from the mean density $(\delta_b,\delta_d)=(0.0,0.0)$ to higher birth and death densities. Overall, it reveals that towards lower redshifts we see the formation of mass islands and clumps over an increasing range of density values. These features also exist over an order of magnitude higher density range, implied by the increased persistence range. They also merge with surrounding structures at a higher and wider range of positive density values. The development of a wider and richer population of mass clumps is a direct manifestation of the hierarchical nature in which they build up.

\item The triangular core region in the one-dimensional persistence diagrams show a moderate expansion from $z=3.8$ to $z=0.0$. It is widening to both lower and higher densities, including a mild increase of the persistence values. The triangular region is and remains quite symmetric, while its location hardly shifts. Its evolution is mainly one in which the left- and right-hand concave wings -- seen along the birth-death line -- gradually move up and outward. Having noted that the prominent features in the one-dimensional persistence diagram represent the phase in which filaments and tunnels connect the overdense regions in the cosmic mass distribution into the pervasive structure of the cosmic web, its moderate development shows that this transition retains a largely universal character with only a mild change of the densities of the filamentary connections. 
  
\item Interestingly, the evolution of the two-dimensional void persistence diagrams appears to be dominated by a shift in density values, and considerably less by a widening of the density values of the voids. The increase in the density and persistence range of voids is quite limited. Instead, we see a continuous shift from $z=3.8$ to $z=0.0$ of the persistence points to lower density values. It is a direct reflection of the outflow of mass from the void interior and the continuously deepening of the void interior~\citep[see e.g.][]{weygaert1993,shethwey2004}, in combination with the restricted density range of voids to $-1.0<\delta<0.0$. 
\end{itemize}

\bigskip
\noindent In addition to these general observations concerning the evolution of the persistence diagrams, we wish to address two characteristics and/or signatures that in the earlier discussion on the present epoch ($z=0.0$) persistence diagram were identified as providing specific information on the formation of the cosmic web and its connections. The first aspect is the presence of an apex in the persistence diagrams, the second aspect the distribution of the persistence values of topological features. 

\begin{figure*}
 \vspace{1.0truecm}
 \centering
  \includegraphics[width=\textwidth]{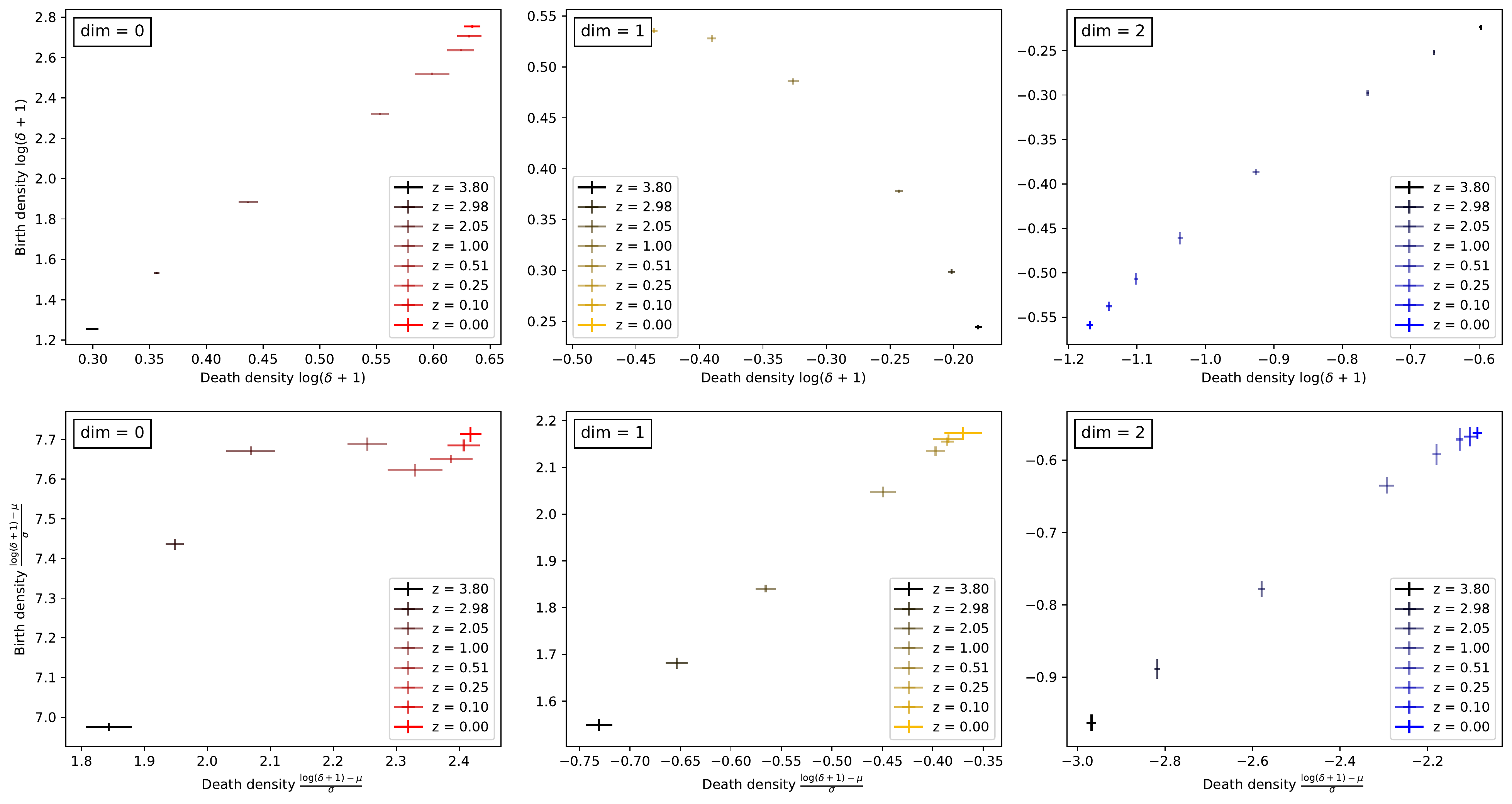}
 \caption{\textbf{Evolution of $\bm{\Lambda}$CDM homology, the apex.} We show the evolution of the high-persistence apex for the usual density field and for a field scaled so that $\sigma_{\log(\delta+1)}=1$ at all redshifts, thus reducing the influence of evolved structures. The error bars show the standard deviation of the location's birth and death value as calculated from the five independent simulation runs.}\label{fig:13_apex_evolution}
  \centering
  \includegraphics[width=\textwidth]{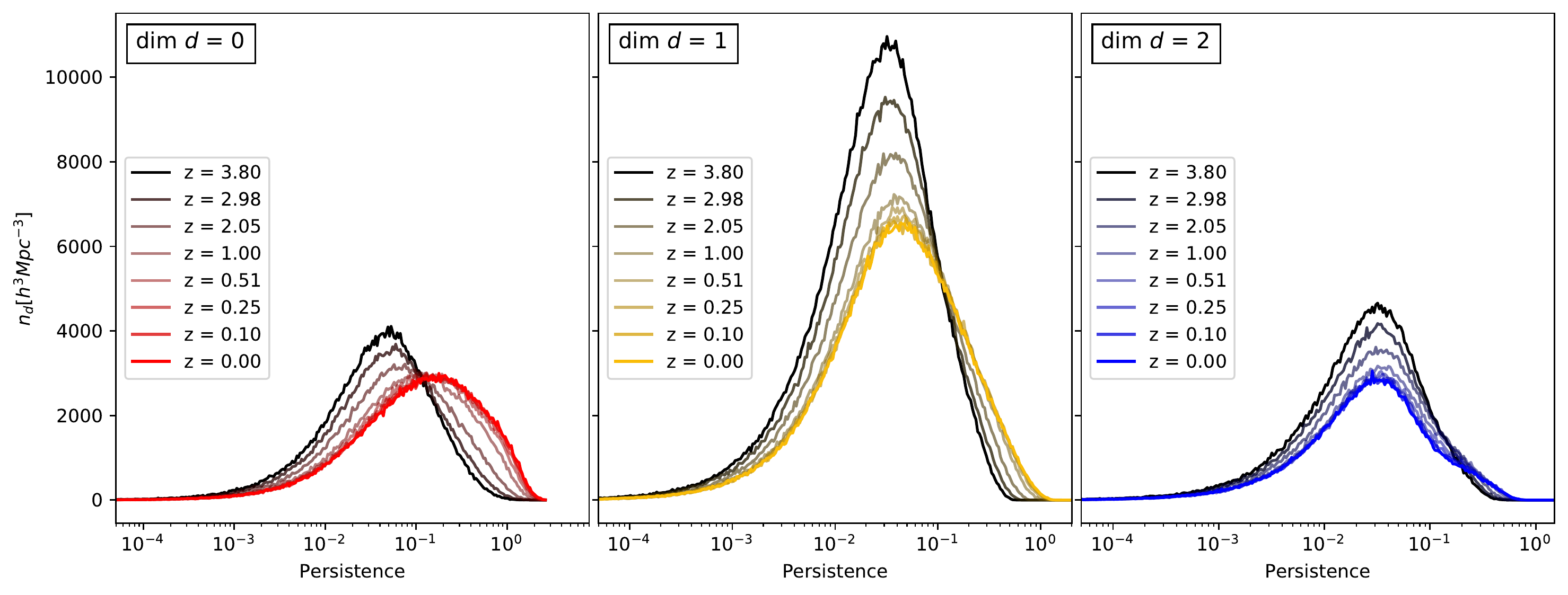}
 \caption{\textbf{Evolution of $\bm{\Lambda}$CDM homology, Persistence curves.} Persistence curves of dimensions zero, one and two (left to right), with the curves for all eight different redshifts together. To illustrate the time evolution, the curve corresponding to the earliest snapshot (redshift $z = 3.8$) is the darkest (black) and brightness increases when progressing towards lower redshift, until the curve for redshift $z = 0$ is red, yellow or blue, according to the respective dimension of zero, one and two. We see a clear decrease in the number of features at the lower-persistence side, and a consistent increase at values of higher persistence.}\label{fig:14_persistence_distribution}
\end{figure*}

\subsubsection{Evolving persistence and connectivity: the apex transition}
\label{sec:apex}
The multiscale nature of the gravitationally evolved mass distribution at $z=0.0$ is marked by the presence of a distinct apex in the persistence diagram (Section~\ref{sec:persistence_web}). The sharp apexes in the zero-, one- and two-dimensional diagrams turn out to be manifestations of a characteristic transition in the dynamical structure and development of the cosmic mass distribution. The apex in the zero-dimensional diagram marks the overdense features that are connecting up into the pervasive network of the cosmic web. The connection typically occurs at the density level at which these features turn around their initial expansion into gravitational contraction. This important connectivity transition is also recognized as an apex in the one-dimensional diagram marking the birth of the filaments and tunnels that form the bridges of the cosmic web. The apex in the two-dimensional persistence diagram for voids signifies the hierarchical evolution of the void population, marking the density at which they emerge as enclosed cavities and also the characteristic density $\delta \approx -0.8$ of fully evolved voids. 

Given the significance of the apexes for our understanding of how the various topological features connect up in the cosmic web and of their role in establishing these connections, we have assessed how their locations in the persistence diagrams evolve in time. To this end we determine the tip of the apexes in the zero-, one- and two-dimensional persistence diagrams in terms of the birth and death density values, and take this location as defining parameter for the apex. Concretely, we computed this by taking the mean of all birth-death pairs with a persistence higher than 99.9\% of the highest occurring persistence (for this dimension). The evolution of the apexes in the three persistence diagrams is shown in Fig.~\ref{fig:13_apex_evolution}. The top row of the figure shows the evolution of the apex's location in terms of birth density $\delta_b$ and death density $\delta_d$. The uncertainties are calculated from the five independent simulation runs.

The trends in terms of the regular (birth,death) values are as expected. First, we observe a generic cosmological aspect in all persistence diagrams. In all panels we see that the apexes experience clear and uniform evolutionary trends. Also, we see that the increase or decrease of the apex locations appears to slow down after $z=1.0$. The changes in apex location from $z=1.0$ to $z=0.0$ are hardly significant. This is a direct reflection of the slowdown in structure growth in \lcdm cosmologies due to the increasing dominance of the cosmological constant. In the second publication of this study~\citep{wilding2021}, we address the aspect of the global cosmological and dark energy influence on the topology of the cosmic web in considerably more detail. 

A second global observation is that all apexes in the zero-, one- and two-dimensional persistence diagrams experience a strong, uniform shift through the persistence diagrams. The development in each dimension exhibits some notable differences. The apex in the zero-dimensional diagram for mass concentrations shows a steady and uniform increase. On the other hand, the apex in the one-dimensional diagram for filamentary loops moves towards higher birth densities, while the density at which they merge into walls surrounding voids decreases with time. Finally, the two-dimensional apex, marking the appearance of voids and their disappearance into the overall mass distribution, shows a uniform decrease to ever lower density values as time proceeds.

The apex in the zero-dimensional persistence diagram (Fig.~\ref{fig:13_apex_evolution}, left-hand top row) shows a uniform increase towards decreasing redshift. As time proceeds, this reflects the growing density contrast of the matter concentrations. Topologically speaking, this results in the appearance of overdense islands at higher birth densities, as well as their merging with nearby features and the cosmic web at correspondingly higher densities. This time evolution proceeds by the hierarchical buildup of the mass concentrations: the peaks and islands at higher redshift were of a smaller spatial scale, and gradually merge into more massive and larger objects. The smaller density clumps at higher redshifts correspond to lower densities (at the effective gridcell filter scale).

The most interesting issue with respect to the formation of the cosmic web is how the apex in the one-dimensional persistence diagram of filaments and tunnels develops as mass concentrations condense out of the primordial density field. From the top centre panel of Fig.~\ref{fig:13_apex_evolution} we learn that the birth density at which filaments and tunnels appear in the density field grows in time. As with the mass clumps, this is a direct reflection of the hierarchical buildup of all aspects of the cosmic web. At higher redshifts, the filamentary bridges have a smaller scale. Even when compensating for the overall growth of structure in terms of the normalized density (Fig.~\ref{fig:13_apex_evolution}, bottom centre panel), we find a continuous increase of the birth density of filaments and tunnels as time proceeds. This is the result of the buildup of filaments through a process of continuous merging of smaller filamentary bridges into larger scale filaments. Visually, we see this process in computer simulations, such as in Fig.~\ref{fig:1_gadgetdtfe}.

Descending to even lower density thresholds we see how filaments disappear as they blend into the wall-like boundaries surrounding void cavities. As time proceeds, we see this happening at ever lower density thresholds (Fig.~\ref{fig:13_apex_evolution}, top centre panel). It reflects the fact that voids evacuate their interior as they mature, resulting in a mass distribution marked by ever larger and emptier voids. The void population builds up in a hierarchical fashion, in which large voids are the product of the merging of smaller voids that dominated the mass distribution at earlier epochs~\citep{shethwey2004}. The self-similar development seen in the two-dimensional persistence diagram of normalized densities (Fig.~\ref{fig:13_apex_evolution}, bottom right-hand panel) is a direct reflection of this.

\subsubsection{Evolving persistence and connectivity: self-similarity?}\label{sec:selfsimilar}
We assess the location of the apexes with respect to the overall evolving mass distribution. Given the non-linear nature of the evolved mass distribution at the redshifts analysed, we do this in terms of the normalized (logarithmic) density $f_n$, 
\begin{equation}
  \label{eq:f-sigma}
  \bar{f_n}=\frac{f_l - \mu_l}{\sigma_l} \,,
\end{equation}
\noindent in which $f_l$ is the logarithmic value of the density field, 
\begin{equation}
f_l\, = \, \log(\delta +1)\,,
\end{equation}
\noindent and $\mu_l$ and $\sigma_l$ its mean and dispersion. In a hierarchically evolving mass distribution, the structure on small scales at earlier times resembles that of the structure on large scales at later times. Ideally, in scenarios in which the primordial mass distribution would be described by a power-law power spectrum, the resemblance would be one of perfect self-similarity: the mass distribution at early times would be a small-scale version (statistically) exactly similar to the large-scale distribution at later times. The Megaparsec cosmic web at $z=0.0$ is similar to the cosmic web that existed on smaller scales at higher redshifts. While a \lcdm power spectrum does not result in a perfect self-similar evolution, over the spatial scales covered by the simulation box we should to good approximation expect self-similar behaviour. 

As we discussed in Section~\ref{sec:apex}, the apexes are topological signatures of dynamical transitions in the buildup of the cosmic web. In a perfectly self-similar cosmology, at each scale these transitions would occur under the same conditions, at correspondingly different epochs. At earlier epochs the apexes, and the transitions they entail, should also occur at similar densities on smaller scales. To this end, we compare the density values of the apexes in normalized units, i.e., in terms of $\bar{f_n}$ (equation~\ref{eq:f-sigma}).

By scaling the mass distribution at each epoch in terms of the overall amplitude of the density inhomogeneities, we test whether, in essence, its evolution is a self-similar mapping from epoch to epoch. The panels in the bottom row of Fig.~\ref{fig:13_apex_evolution} plot the location of the apexes for the zero-, one- and two-dimensional persistence diagrams in terms of the scaled mass distribution. It reveals that at redshifts from $z \approx 1.0$ to $z=0.0$, the apex is found at approximately the same birth and death value (within error bars). All three dimensions also show a uniform increase of the apex (birth,death) values from high redshift to low redshift, which is as expected given the considerably smaller characteristic scales of the cosmic web at higher redshift given the effective filtering on a gridscale. 

The normalized persistence diagrams show a different behaviour with respect to their apexes: for all three diagrams we find a systematic and uniform increase of the apexes' normalized density from high to low redshift. It is strong evidence for the persistence diagrams topologically expressing the notion of the self-similar evolution of the mass distribution.


\subsubsection{Evolution of persistence values}
An additional informative aspect are the persistence values for the topological features in the mass distribution. Fig.~\ref{fig:14_persistence_distribution} shows the evolving distribution of persistence values for the zero-, one- and two-dimensional features in the mass distribution. The panels plot the number density of features as a function of persistence value $\pi_i$. The time evolution is represented in terms of the changing colour of the curves, turning from dark at $z=3.8$ to red, yellow or blue at $z=0.0$. The persistence values have a range spanning several magnitudes, hence the logarithmic scale for the persistence values.

For all three dimensions, the persistence distribution is a distribution that peaks at a characteristics persistence, with a value between $\pi \approx 0.05$ and $\pi \approx 0.1$. On the left-hand side this is preceded, in all cases, by a long wing of low persistence values, in essence the noisy features in the mass distribution. The right-hand wings represent the stable and prominent features that exist over a large density range. 

The evolution of the persistence curves involves two aspects. The first one is the uniform decrease of the curves on the low persistence side from $z=3.8$ to $z=0.0$. It reflects the gradual disappearance of noisy feature. We observe this in the case of the mass clumps, of the filaments and tunnels an in the case of the voids.

The second aspect of the persistence evolution reveals a difference between the three classes of features. There is a strong increase of high persistence mass clumps (zero-dimensional features). This is immediately attested by the systematic and sizeable shift of the high persistence wing from $z=3.8$ to $z=0.0$. It even involves a shift of the peak of the distribution towards higher persistence values. At later times, gravitational evolution has produced highly non-linear massive clumps that mark more prominent and stable features in the overall mass distribution than their less pronounced precursors. 

Also, the persistence values of the one-dimensional features (filaments and tunnels) show a trend towards higher persistence values. The trend is more moderate than that for the zero-dimensional features, and also involves only a minor shift of the peak towards higher persistence values. The filaments and tunnels appear to become a more robust element of the weblike network in which the mass distribution is organized. It is a reflection of the hierarchical evolution of the cosmic web, in which more tenuous filaments and tunnels define a smaller-scale version of the cosmic web at high redshifts, while larger, denser filamentary bridges mark the mass distribution at later times~\citep[see e.g.][for an extensive description]{cautun2014}.

The voids hardly reveal a shift towards higher density values. Their evolution predominantly involves an almost uniform decrease in number density of voids of a given persistence. The slight but still perceptible increase in the number of high persistence features (albeit much smaller than in the lower dimensions) points to the formation of a small number of large, deep voids. The evolution also reflects the almost self-similar development of the hierarchically evolving void population: the late-time void population is -- statistically speaking -- a large-scale equivalent of the population of smaller voids at earlier epochs~\citep{shethwey2004}.

%% file: 5_summary.tex
\section{Summary and conclusions}\label{sec:summary}
We assess the topological structure and connectivity of the \lcdm cosmic web in terms of the multiscale topological formalism of persistence and Betti numbers. TDA offers an intricate quantitative description of how the structural components of the cosmic web are assembled and organised within its complex network. The Betti curves specify the prominence of features as a function of density level, and their evolution with cosmic epoch reflects the changing network connections between these structural features. The persistence diagrams quantify the longevity and stability of topological features. In the present study we establish, for the first time, the link between persistence diagrams, the features they show, and the gravitationally driven cosmic structure formation process. By following the diagrams' development over cosmic time, the link between the multiscale topology of the cosmic web and the hierarchical buildup of cosmic structure is established.

Persistent topology enables us to explore the cosmic web's complex and intricate spatial pattern, by specifically addressing the aspects of patterns, connectivity and complexity that are not or hard to infer from the more conventional clustering measures such as the two-point correlation functions. In this sense, persistent homology provides us with necessary complementary phase correlation information on the large-scale distribution of mass and galaxies. It provides an innovative path towards opening up the cosmological information contained in the properties of the cosmic web.

In the present study we describe a detailed and extensive analysis of the evolving hierarchical topology of the cosmic web in \lcdm cosmologies on the basis of the mass density field filtration. The principal intentions are
\begin{itemize}
\item to assess and quantify
the connectivity of the cosmic web in terms of the levels at which its structural components join into the overall weblike network,
\item to establish the relationship between the characteristics of the Betti curves and persistence diagrams, and the gravitationally driven cosmic structure formation process, 
\item to explore the sensitivity of the topology of the cosmic web to the underlying cosmology
\item to assess the extent to which the topological measures are able to extract cosmological information.
\end{itemize}
The present study extends the earlier work by our group on the homology and persistent topology of the cosmic mass distribution~\citep{weygaert2011alpha,nevenzeel2013triangulating,park2013,pranav2017topology,pranav2019unexpected,pranav2019topology,feldbrugge2019} and focuses on the topology of the evolving non-linear cosmic mass distribution in \lcdm cosmologies. 

\subsection{Cosmic web at z=0 -- global and multiscale topology}
The first stage of our analysis is an in-depth investigation of the cosmic web topology at $z=0$, the present epoch. We analyse the Betti curves for zero-, one- and two-dimensional topological features along with the corresponding persistence diagrams. In the physical context of the cosmic web, the zero-dimensional features are the matter assemblies or ``islands'', the superclusters and clusters in the galaxy distribution. The one-dimensional features are the filaments and tunnels, while the two-dimensional features are the low-density voids and their wall-like boundaries. 

\begin{itemize}
\item All three persistence diagrams have a characteristic triangular shaped morphology. The majority of birth-death points are located near the diagonal base, corresponding to low-significance short-lived features. The structurally significant part of the diagrams is the typically triangular region, bounded by the diagonal and two concave edges, which coalesce at a sharp peak.
\item We introduce the concept of the \emph{apex} of a persistence diagram. It refers to the tip of the sharp peaks in the diagrams and represents features of highest persistence. The apex marks the location in the diagram where a large number of features simultaneously undergoes a topological ``phase transition'', either over a wide range of birth or death densities.
\item The formation of the cosmic web is marked by the apex of the one-dimensional persistence diagram. At a birth density of $\delta \approx 5$, it corresponds to the rather sharp transition at which individual mass concentrations -- superclusters and clusters -- get connected through filamentary bridges, establishing the percolating network of the cosmic web.
\item Interestingly, the topological transition at $\delta \approx 5$ coincides with the density at which overdense regions decouple from the Hubble expansion. It indicates an interesting concordance, within a very narrow density range, between the condensation of mass concentrations in the universe and their assembly into a space-filling filamentary network.
\item The disappearance of the filamentary network at a narrow density range around $\delta\approx -0.7$ identifies the stage at which the cosmic mass distribution gets marked by the appearance of a population of individual voids surrounded by enclosing walls. The sharp transition is preceded by a stage in which tunnels get filled up into tenuous solid walls. This transition and the establishment of the prominent void population at $\delta\approx -0.8$, as expected by theories of void evolution~\citep{blumenthal1992,shethwey2004}, is most clearly reflected in the two-dimensional persistence diagram in terms of a sharply outlined apex.
\end{itemize}

\subsection{Cosmic Web evolution -- a dynamic topology}
In the second stage of our analysis we follow the evolving topology of the cosmic web from $z=3.8$ until the current epoch.

\begin{itemize}
\item The overall development of the structure and topology of the cosmic mass distribution is reflected in the change of the corresponding Betti curves. Their parametrization in terms of a skew normal distribution allows a quantitative characterization of the changing properties of the various topological features.
\item A typical example is seen in the evolution of the void population. The outflow of mass from cosmic voids into the surrounding walls and filaments and the corresponding decrease of void densities finds its topological expression in the downward trend of the mean skewed normal parameter $\mu$ (Fig.~\ref{fig:9_fit_parameter_evolution}) of the $d=2$ Betti curve. 
\item The apex of the persistence diagrams displays a systematic shift (see Fig.~\ref{fig:13_apex_evolution}) that reflects the evolution of the structural components of the cosmic web. In terms of the evolving amplitude of the density fluctuations in the cosmic density field, we find that in particular the one-dimensional and two-dimensional apexes show a self-similar development. It demonstrates that the hierarchical nature of the evolution of the filamentary network and void population leaves a topological imprint in the persistence diagram.
\item In the case of all structural components, the systematic shift of the apex slows down considerably after $z\approx0.5$. It corresponds to the slowing and halting of the cosmic structure formation process once dark energy assumes dominance over the dynamics of the universe~\citep{peebles1980,frieman2008}.
\end{itemize}

\subsection{Future outlook}
We have demonstrated that it is possible to obtain a wealth of detailed information on the formation and evolution of large-scale cosmic structure, and specifically the cosmic web, through the analysis of the topological characteristics and connectivity. We found that it is possible to infer the relationship between the abstract language of homology and algebraic topology and a range of aspects of the gravitational structure formation process.

In general, topological characteristics are not related to a single, unique and identifiable structural component. They inform us about the connections with other structures, and hence addresses their global embedding and connectivity. We exploit this in a follow-up study~\citep{wilding2021}, in which we assess the cosmological sensitivity of persistence based topological characteristics. In this study we demonstrate the way in which different dark energy prescriptions translate into detectable and significant differences in topology.

Of key importance for the viability of the topological analysis of the cosmic mass and galaxy distribution is its success with respect to the observational reality. On the one hand, the analysis of observational data poses a range of practical challenges. These include measurement uncertainties, under-sampled regions, regions with non-existing or simply missing data, systematic selection effects and the influence of redshift distortions in galaxy survey maps~\citep[see][]{kaiser1987,hamilton1998}. Investigation of these observational influence is currently the focus of an extensive project, which will be the reported in a future publication~\citep{wilding2021b}.

On the other hand, we need a method that doesn't solely rely on local data, but rather on the interplay between the local and the global structure: The Betti numbers and the detailed persistence persistence diagrams provide exactly that. Research on using persistence to investigate structural patterns in the halo distribution of the cosmic web and its relation to the underlying topology is reported in an accompanying third publication~\citep{bermejo2020}.